\RequirePackage[l2tabu, orthodox]{nag}

\documentclass[fontsize=12pt,a4paper,numbers=enddot,parskip=half]{scrartcl} 

\usepackage{float}

\usepackage[T1]{fontenc}
\usepackage{lmodern} 

\usepackage{amsfonts}
\usepackage{amsmath}
\usepackage{amsthm}
\usepackage{amssymb}
\usepackage{eurosym}
\usepackage{bbm}

\setkomafont{section}{\rmfamily\fontseries{b}\selectfont}
\setkomafont{subsection}{\rmfamily\itshape\mdseries}
\setkomafont{subsubsection}{\rmfamily\itshape\mdseries}
\setkomafont{paragraph}{\rmfamily\itshape\mdseries}
\setkomafont{title}{\rmfamily\large\fontseries{b}\selectfont}
\setkomafont{captionlabel}{\rmfamily\large\fontseries{b}\small\selectfont}

\usepackage{multirow}
\usepackage[font={small},justification=raggedright,singlelinecheck=false]{caption}
\newcommand{\mytablefont}{\vspace{2mm}\fontsize{9}{10.8} \selectfont}

\usepackage{tikz}

\usetikzlibrary{positioning,arrows}

\usepackage[caption=false]{subfig}
\usepackage{todonotes}

\usepackage[ansinew]{inputenc}
\usepackage[english]{babel}

\begin{hyphenrules}{english}
\end{hyphenrules}

\usepackage[top=3cm, bottom=3cm, left=3cm, right=3cm]{geometry}
\usepackage{setspace}
\usepackage{rotating} 
\usepackage{dcolumn}
\newcolumntype{d}[1]{D{.}{.}{#1}}

\usepackage{graphicx}
\usepackage{booktabs}
\usepackage{epstopdf}
\usepackage{enumerate}
\usepackage[hang]{footmisc}
\usepackage[longnamesfirst]{natbib}
\usepackage{authblk}
\usepackage{colortbl}
\usepackage{placeins}
\usepackage{tabularx}
\usepackage{makecell}
\usepackage{wrapfig}
\newcolumntype{Y}{>{\centering\arraybackslash}X}
\newcolumntype{R}{>{\flushright\arraybackslash}X}
\newcolumntype{L}[1]{>{\raggedright\let\newline\\\arraybackslash\hspace{0pt}}m{#1}}
\renewcommand{\arraystretch}{1.2}

\author{
\vspace{5mm} \small MORITZ SCHERRMANN\thanks{Institute for Finance \& Banking, Ludwig-Maximilians-Universit\"at M\"unchen, Ludwigstr.\ 28 RB, 80539 Munich, Germany. E-mail: scherrmann@lmu.de}}
\title{\Large Multi-Label Topic Model for Financial Textual Data}
\date{}

\begin{document}

\maketitle
\vspace{2cm}
{\rmfamily\fontseries{b}\selectfont Abstract}

\smallskip
\small \singlespacing This paper presents a multi-label topic model for financial texts like ad-hoc announcements, 8-K filings, finance related news or annual reports.\\
I train the model on a new financial multi-label database consisting of 3,044 German ad-hoc announcements that are labeled manually using 20 predefined, economically motivated topics. The best model achieves a macro F1 score of more than 85\%. Translating the data results in an English version of the model with similar performance. As application of the model, I investigate differences in stock market reactions across topics. I find evidence for strong positive or negative market reactions for some topics, like announcements of new \textit{Large Scale Projects} or \textit{Bankruptcy Filings}, while I do not observe significant price effects for some other topics. Furthermore, in contrast to previous studies, the multi-label structure of the model allows to analyze the effects of co-occurring topics on stock market reactions. For many cases, the reaction to a specific topic depends heavily on the co-occurrence with other topics. For example, if allocated capital from a Seasoned Equity Offering (\textit{SEO}) is used for restructuring a company in the course of a \textit{Bankruptcy Proceeding}, the market reacts positively on average. However, if that capital is used for covering unexpected, additional costs from the development of new drugs, the \textit{SEO} implies negative reactions on average.
\normalsize
\vspace{2cm}

\newpage
\onehalfspacing

\section{Introduction}

The analysis of the impact of news on financial markets has been a crucial topic in finance for many years. The efficient market hypothesis, proposed by \cite{fama1970}, suggests that all publicly available information is rapidly reflected in stock prices. However, finding explanations of which information in financial news drives stock prices in certain directions may be more complex. Several studies try to explain stock market reactions on corporate news by investigating the tone of a text, for example \cite{loughran}, \cite{malo2014good} or \cite{sinha2022sentfin}. An alternative methodology involves to investigate these reactions by differentiating among various topics within the news, as for example \cite{antweiler2006us}, \cite{neuhierl2013market} or \cite{feuerriegel2016analysis}. 

Building upon the existing literature, this work delves deeper into the nuances of market reactions to news topics. The contributions of this study are threefold. First, it introduces a manually-labeled, finance-specific German textual data set, filling a notable gap in the existing literature. Second, the research involves the training of a state-of-the-art German language model specifically tailored for finance-related topic classification. Lastly, the model is designed to be capable of multi-label predictions, thereby accommodating a range of scenarios from texts that focus on a single financial topic to those that span multiple distinct topics. Even predictions with no topic label at all are possible, for texts that cover content not related to finance. This is an extension to the existing topic model literature, which exclusively allows for one topic per document. When linked to stock market reactions, the model provides insights into typical price movements when company news contains a specific topic or even combinations of topics. When a financial news contains more than one topic, the model helps to unfold the different, sometimes even opposite effects of co-occurring topics on the overall abnormal return after an announcement. The multi-label structure additionally allows to investigate how the effect of a specific topic on price movements changes if the respective topic appears in the context of other topics. The model can be applied to documents of arbitrary lengths covering an arbitrary number of topics, as the model predicts topics sentence by sentence and aggregates them to topic labels on document level. This helps to circumvent the limited input lengths of typical language models, so that no relevant information will be dropped. By translating the manually-labeled data set to English, I am able to provide models for both German and English texts.

The model presents multiple applications. Investors could employ the model to categorize news pertaining to a particular firm by topics, thereby gaining insights into the firm's recent activities or overall financial standing. Additionally, they could leverage observed stock market reactions to specific topics as a heuristic for investment decisions following a company announcement. On the corporate side, the model could serve as an auxiliary tool for evaluating whether undisclosed information could be market-relevant, thereby necessitating immediate disclosure in compliance with legal requirements.

To train the multi-label topic model, I use 3,044 German ad-hoc announcements that are manually labeled sentence by sentence by nine financial experts into 20 predefined, economically motivated topics. I conduct several performance and inter-annotator agreement measures to quantify the labeling quality. According to \cite{landis1977application}, the resulting Fleiss' $\kappa$ values of 69.1\% on sentence level and 74.6\% on document level imply a substantial agreement between the annotators. The high average F1 scores among annotators with respect to the labels of the instructors confirm that finding (76.2\% \& 82.2\%, respectively). The labeled sample is called the \textit{Ad-Hoc Multi-Label Database}.

I train the topic model, which is based on the BERT-base model with a classification layer on top, on the new ad-hoc multi-label database, together with three benchmarks. The BERT model outperforms all benchmarks with a macro F1 score of 85.3\% by at least 7.1 percentage points. The BERT performance is highly volatile among topics: Between the best topic (\textit{Squeeze out}) with an F1 score of 96.2\% and the worst topic \textit{Profit Warning} with an F1 score of 67\%, there is a gap of almost 30 percentage points.

As an application I investigate the impact of the topics of an ad-hoc announcement on the respective stock market reactions. The results suggest that there are substantial differences in the typical stock market reactions across topics. There are topics that typically induce very positive market reactions, as for example \textit{Large Scales Projects}. On the other hand, topics like \textit{Bankruptcy Filings} generally lead to very negative market reactions. Some topics, for example \textit{Split}, do not induce any significant market reaction at all. Furthermore, I find evidence that stock market reactions on topics depend on the co-occurring topics. For example, if the capital that is allocated with a Seasoned Equity Offering (\textit{SEO}) is used for restructuring a company after a \textit{Bankruptcy Filing}, the market reacts positively on average. If that capital is used for covering unexpected, additional costs of the development of new drugs, this implies negative reactions on average. Another example is the \textit{Bankruptcy Filing} topic: If it occurs exclusively in an ad-hoc announcement, abnormal returns decrease by 16.02 percentage points. This effect reduces to less than 4 percentage points if the topic co-occurs with a \textit{Bankruptcy Proceedings}, as this implies that the information about the bankruptcy filing is more likely to be already known by the market. These results highlight the benefit of the multi-label approach of this paper.

The remainder of this paper is organized as follows: Section \ref{paper1_sec_literature_review} introduces the existing literature about topic modeling in finance and highlights the differences to my model. Section \ref{paper1_sec_mutliLabelDatabase} provides all the steps for the preparation of the ad-hoc multi-label database. Section \ref{paper1_adHocClassification} describes the used topic model, its training algorithm together with the benchmarks and compares their out-of-sample performances on the ad-hoc multi-label database. Section \ref{paper1_sec_StockMarketReaction} is an application of the topic model which classifies all available ad-hoc announcements into topics and links them to their respective stock market reactions in order to investigate typical price reactions to topics. Section \ref{paper1_sec_conclusion} concludes and provides limitations of the model.

\section{Literature Review}
\label{paper1_sec_literature_review}
The following section provides an overview of the research about the categorization of financial news. The first study that covered that field of research is the work of \cite{antweiler2006us}, where the authors inspect the validity of the efficient market hypothesis \citep{fama1970} by observing the stock-market reactions on more than 250,000 Wall Street Journal corporate news stories from 1973 to 2001. To do so, they define 43 topics to which they manually allocate 2,000 randomly picked news. They fit a Na\"ive Bayes classifier that is able to assign the remaining news to the defined topics. They find that the efficient market hypothesis is only partly correct. The typical response to a news story is a strong and prompt reaction followed by a gradual and lengthy reversal. The direction of the initial response is dependent on the topic, for example \textit{Earnings Up} and \textit{Earnings Down}.

\cite{neuhierl2013market} manually classify a data set of 271,867 US corporate press releases between 2006 and 2009 by topic and examine the market response to different types of news. They investigate the impact of various types of corporate announcements on stock returns, volatility, bid-ask spreads and trading volume using these measures as metrics for the informative value of the news. The authors define 10 major news categories that are further subdivided into 60 subcategories. They find that most types of press releases lead to a decrease in the level of informational asymmetry in the market. Furthermore, \cite{neuhierl2013market} find that volatility tends to increase following most types of announcements.

\cite{boudoukh2013news} use a rule-based information extraction tool called \textit{The Stock Sonar (TSS)} that is able to extract the sentiment and the event category of a news. With that tool at hand, they are able to classify all news from the Dow Jones Newswire between 2000 and 2009 in 14 event categories with 56 subcategories. The authors investigate the key features of financial news that drive stock prices. They find that there is a close link between stock prices and information when information about the news like the topic and the tone are taken into account.

The study of \cite{feuerriegel2016analysis} applies an unsupervised topic modeling approach with \textit{Latent Dirichlet Allocation (LDA)} \citep{blei2003latent}. Their aim is to analyze the effect of underlying topics in German financial news on stock prices. The sample consists of 7,645 regulated ad-hoc announcements gathered from the EQS news group between 2004 and 2012, allocated to 40 topics found by the LDA model. The authors find great differences in the stock price effects between topic groups. 

The study of \cite{feuerriegel2021investor} is very similar to the one of \cite{feuerriegel2016analysis} as they use the exact same approach, which is the topic modeling of financial announcements using LDA. The only difference is the application on US data, as their sample consists of 73,986 regulated 8-K filings from companies listed on the New York Stock Exchange between 2004 and 2013. The authors are able to identify 20 different topics using their LDA model. Also for US news, the authors determine a  discrepancy among various types of news stories with respect to their impact on financial markets.

This study differs from the mentioned research in several aspects.
First, all of the mentioned studies consider the annotation of financial news as a multi-class problem, which means that every news can only be assigned to exactly one of N possible topics. However, many news cover different topics at the same time, as for example \textit{Earnings} announcements often co-occur with a \textit{Guidance} about future profit expectations. If that is the case, it is hardly possible to map the effect of one single topic to, for example, stock prices, as there might be a latent topic which affects stock prices as well. This is why I propose a new approach which allows news to have more than one topic. In other words, I define the annotation process as a multi-label problem. This approach allows for example to disentangle which of the underlying topics within one announcement is the key driver for stock price effects.\\
Secondly, the database employed in this study is annotated on sentence level, in contrast to other studies that utilize document-level labeling. Most of the language models are only able to process texts of a limited number of tokens, or their performance strongly decreases with increasing input lengths. Many models automatically truncate inputs to a specific length. For example, BERT models of \cite{bert} allow inputs with a maximum of 512 tokens. Many financial news or announcements exceed this length by far, so that it might happen that relevant information, appearing in the middle or at the end of some news, will never be processed by the model. In contrast, the incidence of sentences surpassing the typical maximum input size is rare, thereby ensuring that a sentence-level labeling approach results in minimal information loss.\\
Third, none of the mentioned studies discuss or review the quality of their topic labels. For manually labeled data sets, as it is the case for \cite{antweiler2006us} and \cite{neuhierl2013market}, no details about the annotation process, the annotation rules or the inter-annotator agreement are given. Even the automatically generated labels in the studies of \cite{boudoukh2013news}, \cite{feuerriegel2016analysis} and \cite{feuerriegel2021investor} miss any form of validation, for example through the manual review of a subset of the samples. Furthermore, the results obtained from methodologies such as LDA exhibit a strong dependency on hyperparameter selections, particularly the predetermined number of topics, and have been subject to criticism due to their lack of result stability \citep{mantyla2018measuring}. In contrast, I explain in detail the whole annotation process and provide several metrics that measure annotator performance and agreement.\\
Finally, in case a study manually labeled only a subset of their sample, the models used to classify the remaining news in the sample are not up to date, as is the case for the Na\"ive Bayes classifier of \cite{antweiler2006us}. The Na\"ive Bayes assumption states that words in a text are independent of each other. However, in practical applications, words appear in context and thus tend to be highly correlated with each other. More advanced models are able to preserve the structure of a text and are even able to consider the context of words. One example is the BERT model which I use in this study.

\section{Ad-Hoc Multi-Label Database}
\label{paper1_sec_mutliLabelDatabase}
\subsection{Ad-Hoc Topics}
\label{paper1_sec_categories}
Initially, this study focuses on identifying and extracting the most frequent and pertinent topics from German ad-hoc announcements. Only topics that maintain relevance throughout the entire period and that occur with sufficient frequency are considered. For instance, topics related to the Covid crisis are excluded due to their absence prior to 2020. Similarly, announcements concerning companies withdrawing from specific submarkets are not considered as individual topics, despite their presence throughout the period, due to their infrequent occurrence. Such exclusions mitigate potential issues like inadequate coverage in the final database. Table \ref{paper1_tab_category_def} gives an overview of the identified topics.
\begin{table}
\begin{flushleft}
\caption{Topic Definition Ad-Hoc Multi-Label Database}
 \mytablefont{This table presents all 20 topics of the ad-hoc multi-label database together with a short description.}
\label{paper1_tab_category_def}
\mytablefont
\renewcommand*{\arraystretch}{1.7}
\begin{tabularx}{\textwidth}{m{5cm} *{1}{X}}
\toprule
\textbf{Topic} & \textbf{Description}\\
\midrule
Earnings & Earnings Announcement, regular reporting on quarterly or annual results \\
Seasoned Equity Offering (SEO) & Capital increase/reduction by issuing additional shares \\
Management & Any changes in management (board of directors, supervisory board, etc.) \\
Guidance & A company's forecast of its own profit or loss in the near future \\
Profit Warning & Surprising deterioration in earnings/earnings forecast\\
M\&A & New/expansion investment in company or own investment in other company,
incl. acquisition \\
Dividend & Announcement dividend/dividend amount (incl. corrections and expectations) \\
Restructuring & Restructuring measures (processes, organization, capital structure, e.g.: debt-equity
swap, operational restructuring, etc.) \\
Debt & Company issues/returns loan/bond \\
Law & Company is involved in litigation, court case/investigation (case opened/closed,
litigation accruals, sued)\\
Large Scale Project & Completion of major project/order for the company\\
Squeeze Out & Majority shareholder applies for squeeze (transfer of shares held by minority
shareholders to majority shareholder), incl. progress of proceedings\\
Bankruptcy Filing & Company or third party has filed/will file for bankruptcy\\
Bankruptcy Proceedings & Information about concrete progress of bankruptcy proceedings is published\\
Delay & Mandatory report is postponed or not published at all/does not
take place\\
Split & Company carries out stock split\\
Pharma Good & Drug approval/announcement/study success\\
Buyback & Repurchase of own shares\\
Real Invest  & Buying or selling assets such as land, factories, machinery, etc.\\
Delisting & Permanent removal of a stock from a stock exchange\\
\bottomrule
\end{tabularx}
\end{flushleft}
\end{table}
An announcement might belong to more than one topic. The most common example are ad-hoc announcements that report earnings results first (\textit{Earnings}) and forecasts for a future profit or loss last (\textit{Guidance}). Other examples are \textit{Guidance} \& \textit{Profit Warning} or \textit{SEO} \& \textit{M\&A}. However, for announcements whose topics are not covered by those of Table \ref{paper1_tab_category_def}, no label will be present at all. Therefore, the given data is a multi-label problem, where every announcement may theortically belong to a number of topics between zero and 20. 
\subsection{Data Collection \& Preparation}
According to Article 17 of the regulation (EU) No. 596/2014 of the European Parliament and of the council of 16 April 2014, called market abuse regulation (MAR), every company that has requested or approved admission of their financial instruments to trading on a regulated market has to inform as soon as possible the public of inside information which directly concerns that company. In Germany, these news are called ad-hoc announcements. The publication of announcements is almost exclusively carried out via ad-hoc service providers. In Germany and other German-speaking countries, by far the most engaged ad-hoc service provider is the EQS Group (formerly Deutsche Gesellschaft f\"ur Ad-hoc-Publizit\"at mbH (DGAP))\footnote{https://www.eqs-news.com, formerly https://www.dgap.de}. Section 26 of the German Securities Trading Law (WpHG) additionally requires companies to send ad-hoc announcements to the company register\footnote{https://www.unternehmensregister.de}. Therefore, I work with all available data from both sources, the EQS Group and the company register. I acquire the data using a python web crawler. The date of the data acquisition is the 20th of March 2022. For the EQS Group, I end up with 129,121 ad-hoc announcements between 1st of July 1996 and 20th of March 2022. Regarding the company register, it is only possible to crawl the last ten years. As I started the data acquisition in 2018, my sample from the company register consists of 34,106 announcements between 16th of December 2008 and 20th of March 2022.\\
If an announcement appears in both data sources, I use the one of the EQS Group as the coverage of relevant information is better for the EQS Group. Upon consolidation, the final data set comprises 132,371 ad-hoc announcements, of which over 95\% are sourced from the EQS Group. 
\subsection{Annotation Design}
The annotation process is carried out by nine annotators, including two professors, five doctoral students from the finance area as well as two students pursuing a master's degree in business administration. In order to be able to roughly control the distribution across topics in the final labeled data set, I assign preliminary labels to all announcements using the Okapi BM25 retrieval function \citep{robertson1994some} with manually specified keywords for every topic (see appendix for a list of all keywords). Prior to the actual annotation phase, I hold a detailed session for all annotators explaining the task, the topic definitions and the labeling app I programmed personally to fit the special annotation design. Furthermore, all annotators received a file containing general and topic-specific labeling hints (see appendix). The annotator's task is to assign topic labels for every sentence in their sample. Besides the already mentioned benefit of circumventing limited maximal inputs lengths of language models, there are two further reasons why I use a labeling on sentence level instead of labeling on document level: On the one hand, in case that an announcement belongs to more than one topic, it is clear which part of the announcement belongs to which topic. This property makes it easier for future models to learn patterns for specific topics, since there is no ambiguity within topics. On the other hand, the sentence level design should improve the quality of the annotators labels, as the annotator's attention might decline with increasing announcement lengths. However, I am able to restore labels on document level by aggregation of all document's sentence labels. Only sentences that can be clearly assigned to a topic independently of previous and subsequent sentences should be labeled by the annotators. Nevertheless, the announcement process remains a multi-label problem as I allow sentences to have a number of labels between zero and 20. The labeling app allows the annotators to add comments when they are unsure about a specific label. These cases are reviewed and discussed at a later point in time. Additionally, I introduce the \textit{Irrelevant} label which indicates whether a sentence is not part of the core of the announcement, like disclaimers or general information about the company. This label is useful for developing models that are able to filter irrelevant information from ad-hoc announcements. However, the \textit{Irrelevant} label is special in a sense that it does not point to any topic. It is just a tool to clean the data set. Therefore, it is not possible to label any sentence with the \textit{Irrelevant} label together with some other label.

I divide the annotation process into three phases. For the first phase, I sample three announcements for every of the 20 topics. To do so, I require at least one sentence of an announcement to have the respective BM25 preliminary label. Due to the multi-label nature of the data, some announcements appear twice in the sample. After duplicate removal, I end up with 57 announcements adding up to 490 sentences for the first annotation phase. I carefully label all of the 490 sentences together with a finance professor. Our labels serve as a \textit{gold standard} for the remaining seven annotators, whose task in phase 1 is also to label the same 490 sentences. On the one hand, the purpose of the first annotation phase is to investigate whether the remaining annotators understand the labeling instructions for every topic as intended by the finance professor and myself,~i.e. I examine the annotators performance with respect to the gold standard. On the other hand, I measure the inter-annotator agreement, since the goal is to end up with a data set with consistently annotated sentences. Additionally, the first phase serves as a test run for the labeling app, so that possible bugs can be detected and solved.

The main labeling work is done during the second annotation phase. Every annotator labels 320 ad-hoc announcements, of which 300 are only allocated to the specific annotator. The remaining 20 announcements, consisting of one announcement per preliminary topic, are the same for all annotators and again serve to keep control of the annotators performance with respect to the gold standard labels and the inter-annotator agreement. This time, only my labels are defined as the gold labels. The 300 individual announcements are composed in a way so that the topics within the final data set will be as balanced as possible. To do so, I compute the average number of labeled sentences per announcement for every topic, using the labeled announcements of the first annotation phase. In that way, I am able to compute the number of announcements for each topic that an annotator has to receive to end up with a balanced database. If a topic has a high number of labeled sentences per announcement, the annotator receives only few announcements of that topic and vice versa. Since I roughly require 50 labeled sentences for every topic and annotator, I end up with 300 announcements per annotator. Prior to the start of phase 2, an additional session with all annotators is held to analyze the results of phase 1 to discuss topics with low performance and agreement scores and to remove misunderstandings.

The third annotation phase consists of 31 unique announcements for every annotator. These announcements are preliminary labeled with topics that are underrepresented in the sample after the first two annotation phases. The purpose of the last phase is therefore to improve the balance of the labels in the final data set. The main reason why this step is necessary is that some BM25 preliminary labels are erroneous. Furthermore, some topics correlate strongly, which increases their absolute number of occurrences (e.g. \textit{Earnings} \& \textit{Guidance}).

\begin{figure}[t]
\caption{Announcement Distribution Within Ad-Hoc Multi-Label Database by Year}
{\includegraphics[width=\linewidth]{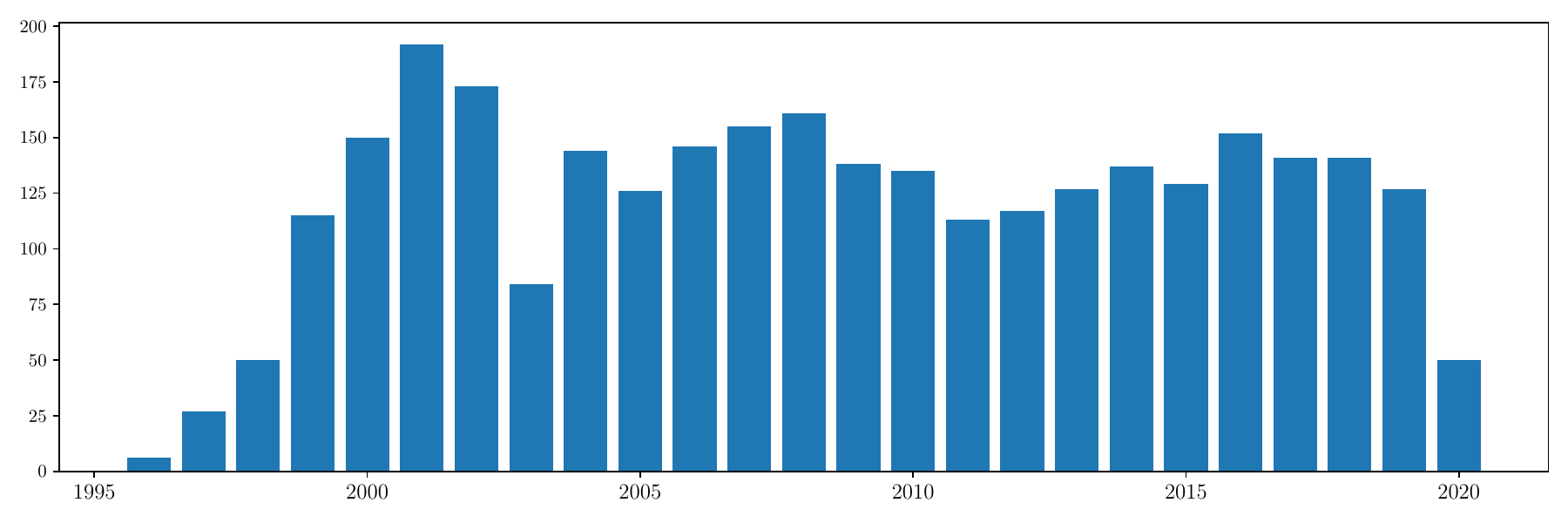}}\par

\mytablefont{A histogram of the distribution of all announcements within the ad-hoc multi-label database by year of the announcement.}
\label{paper1_fig_hist_by_year}
\end{figure}

After duplicate removal, the final sample of the ad-hoc multi-label database consists of 31,771 sentences from 3,044 announcements. Figure \ref{paper1_fig_hist_by_year} shows that the ad-hoc multi-label database consists of news between 1996 and 2020, with a roughly similar number of announcements for all the years. The only exceptions are 2020, 2003 and the first three years. This ensures that the database is not biased for example through over- or underrepresented news during specific periods like the financial crisis in 2008. 
%
\subsection{Data Validation}
The next section describes the annotators' performance with respect to the gold standard and the inter-annotator agreement for both annotation phase 1 and 2.
I compute these measures for both, sentence-level and document-level aggregation. For the sentence-level aggregation, no preparation at all is necessary, as the database is already given on a sentence basis. Document-level aggregation requires the aggregation of all sentences of a document together with their respective labels. The reason for the different aggregation levels is that it is more likely that the labels of two annotators coincide on document level, as there is often more than one sentence that assigns a document to a specific topic. Therefore, it is not necessary that two annotators coincide on all sentences to assign a document to a respective topic; it is enough that one sentence is assigned. However, as mentioned, the drawback is that the document aggregation reduces the number of texts in the sample drastically (31,771 sentences vs. 3,044 documents).
\subsubsection{Annotator Performance Labeling Phase 1}
Table \ref{paper1_prec_rec_f1_annotator_1} measures the performance of the remaining seven annotators with respect to the gold labels. I measure the performance with the precision, recall and F1 score as defined by \cite{sokolova2009systematic} for every annotator and text-aggregation level, averaged over topics. All annotators get a unique but anonymized label A1-A9.
\begin{table}
\begin{flushleft}
\caption{Annotator Performance Metrics in Labeling Phase 1}
 \mytablefont{In labeling phase 1, the gold labels among all common texts are created by two annotators. This table displays macro precision, recall and F1 of the remaining seven annotators (as defined by \cite{sokolova2009systematic}, averaged over topics) as percentage numbers. Additionally, the table displays the number of texts for each annotator. I conduct the analysis on sentence and document level. Annotators are sorted by decreasing F1 score on sentence level.}
\label{paper1_prec_rec_f1_annotator_1}
\mytablefont
\begin{tabularx}{\textwidth}{m{2cm} *{8}{Y}}
\toprule
 \textbf{Annotator} & \textbf{A6} & \textbf{A9} & \textbf{A3} & \textbf{A5} & \textbf{A4} & \textbf{A8} & \textbf{A2} & \textbf{Avg.} \\
\midrule
  \multicolumn{9}{c}{\textit{Panel A: Sentence Level}} \\
       \midrule
Precision & 78.4 & 87.9 & 90.5 & 86.0 & 80.9 & 88.2 & 84.2 & 85.1 \\
Recall & 69.8 & 60.3 & 54.4 & 53.0 & 52.4 & 49.1 & 42.2 & 54.4 \\
F1 & 72.3 & 68.2 & 64.1 & 62.2 & 60.4 & 59.5 & 54.2 & 63.0 \\
Num. & \multicolumn{8}{c}{490} \\
 \midrule 
\multicolumn{9}{c}{\textit{Panel B: Document Level}} \\
       \midrule
Precision & 89.3 & 89.0 & 85.0 & 90.2 & 91.3 & 84.1 & 86.2 & 87.9 \\
Recall & 86.9 & 84.9 & 86.9 & 79.1 & 81.1 & 86.4 & 76.7 & 83.2 \\
F1 & 86.0 & 85.7 & 83.5 & 83.1 & 82.6 & 82.5 & 79.1 & 83.2 \\
Num. & \multicolumn{8}{c}{57} \\
 \bottomrule
\end{tabularx}
\end{flushleft}
\end{table}

Starting with the average annotator performance on sentence level, we see that the average annotators' precision (85.1\%) is clearly higher than the recall (54.4\%). This pattern holds for all annotators, but with different magnitudes. This indicates that when a text is labeled, the label is usually correct (precision), but too many sentences are left without label (recall). All annotators perform similarly with respect to their precision score, which varies between 90.5\% and 78.4\%. However, there are explicit differences among their recall scores, as they vary between 69.8\% and 42.2\%. This results in an overall F1 score of 63\%.

On document level, we observe a benefit of less penalized missing labels, as the gap between the average precision and recall is almost closed (87.9\% vs. 83.2\%). This leads to an increased overall F1 score of 83.2\%. Additionally, the performances between annotators are more balanced. However, this comes at cost of only having 57 observations.

\clearpage
\begin{table}
\begin{flushleft}
\caption{Topic Performance Metrics in Labeling Phase 1}
 \mytablefont{This table displays macro precision, recall and F1 (as defined by \cite{sokolova2009systematic}, averaged over annotators) as percentage numbers for all topics. Additionally, the table displays the number of texts for each topic. I conduct the analysis on sentence and document level. In labeling phase 1, the gold labels among all common texts are created by two annotators. Annotators are sorted by decreasing F1 score on sentence level.}
\label{paper1_prec_rec_f1_category_1}
\mytablefont
\begin{tabularx}{\textwidth}{m{3.5cm} *{8}{Y}}
\toprule
 & \multicolumn{4}{c}{\textit{Sentence Level}} & \multicolumn{4}{c}{\textit{Document Level}} \\
 & Num. & Prec. & Recall & F1 & Num. & Prec. & Recall & F1 \\
\midrule
Management & 6 & 90.7 & 85.7 & 87.3 & 3 & 91.7 & 95.2 & 93.2 \\
Squeeze Out & 4 & 90.0 & 85.7 & 85.5 & 3 & 90.7 & 100 & 94.4 \\
Large Scale Project & 9 & 100 & 66.7 & 80.0 & 3 & 100 & 100 & 100 \\
Split & 10 & 89.3 & 70.0 & 77.7 & 3 & 100 & 100 & 100 \\
Dividend & 20 & 100 & 60.7 & 75.2 & 8 & 100 & 96.4 & 98.0 \\
Delisting & 11 & 88.6 & 63.6 & 71.2 & 3 & 91.7 & 90.5 & 90.3 \\
Earnings & 59 & 90.3 & 61.7 & 71.1 & 9 & 84.0 & 85.7 & 84.5 \\
Guidance & 11 & 60.7 & 70.1 & 64.1 & 7 & 67.7 & 77.6 & 71.5 \\
Pharma Good & 12 & 100 & 46.4 & 62.0 & 3 & 100 & 85.7 & 90.0 \\
M \& A & 15 & 87.6 & 50.5 & 60.9 & 4 & 84.2 & 78.6 & 77.1 \\
SEO & 23 & 90.6 & 47.8 & 59.5 & 6 & 95.9 & 66.7 & 76.5 \\
Delay & 7 & 89.3 & 44.9 & 57.9 & 4 & 100 & 67.9 & 80.3 \\
Profit Warning & 7 & 75.8 & 53.1 & 57.8 & 4 & 82.1 & 60.7 & 67.1 \\
Law & 11 & 96.2 & 40.3 & 54.0 & 4 & 100 & 78.6 & 87.1 \\
Real Invest & 11 & 92.7 & 39.0 & 53.9 & 3 & 92.9 & 95.2 & 93.1 \\
Buyback & 10 & 71.3 & 44.3 & 53.4 & 3 & 78.6 & 76.2 & 74.5 \\
Restructuring & 20 & 73.4 & 41.4 & 51.8 & 6 & 78.6 & 61.9 & 66.7 \\
Bankruptcy Proceedings & 14 & 69.4 & 42.9 & 51.7 & 3 & 65.7 & 76.2 & 69.5 \\
Debt & 23 & 96.3 & 34.2 & 44.9 & 5 & 95.9 & 77.1 & 83.7 \\
Bankruptcy Filing & 5 & 50.7 & 40.0 & 39.5 & 2 & 58.1 & 92.9 & 66.7 \\
 \midrule 
 Avg. & 14 & 85.1 & 54.4 & 63.0 & 4 & 87.9 & 83.2 & 83.2 \\
\bottomrule
\end{tabularx}
\end{flushleft}
\end{table}

Table \ref{paper1_prec_rec_f1_category_1} repeats the analysis of Table \ref{paper1_prec_rec_f1_annotator_1}, but this time averaged over annotators. This helps to understand which topics were problematic for the annotators. On sentence level, we see strong differences between topics. Topics such as \textit{Management}, \textit{Squeeze Out} and \textit{Large Scale Project} have F1 scores above 80\%, whereas \textit{Bankruptcy Proceedings}, \textit{Debt} and \textit{Bankruptcy Filing} have F1 scores below 55\%.

On document level, we observe the overall trend that topics that were problematic on sentence level are also problematic on document level. However, there are exceptions like \textit{Debt}, \textit{Real Invest} or \textit{Law}. These are topics where the gap between precision and recall is especially large. As explained before, for a good performance on document level a high precision score is much more important than a high recall score, as it is likely to find several sentences of a topic in one document. These cases indicate that the annotators generally understand the respective topic definition, but they do not know all the cases and situations that should lead to a label. In other cases, both precision and recall are low. In these cases, the performance is bad on both sentence and document level. This indicates that the annotators have a basic misunderstanding of the topic. An example is the topic \textit{Bankruptcy Filing} with precision scores of 50.7\% and 58.1\% and recall scores of 40\% and 92.9\%, respectively.

Summing up, the main insights of labeling phase 1 are that there is the general tendency of labeling too few sentences and that there are specific topics that might not have been fully understood by the annotators as intended by the instructors. Therefore, I conduct a session after labeling phase 1 with all annotators where I address all the mentioned issues in detail. Furthermore, I explain and discuss problematic topics and give examples of wrong-labeled sentences.
\subsubsection{Annotator Performance Labeling Phase 2} 
For the second labeling phase, I decided to only choose 20 announcements, one for every topic, that are allocated to every annotator for tracking their performance and inter-annotator agreement. The reason for that low number is that I aim to increase the number of uniquely labeled sentences which increases the size of the database while keeping the workload manageable for every annotator. Unfortunately, it turns out that for some topics there are only few or even no gold labels available, which makes the computation of precision, recall and F1 impossible. Therefore, I drop every topic with less than three labeled sentences for the analysis of the annotator performance and inter-annotator agreement. These topics are \textit{Large Scale Project}, \textit{Real Invest},  \textit{Delay},  \textit{Profit Warning}, \textit{SEO} and  \textit{Pharma Good}. Table \ref{paper1_prec_rec_f1_annotator_2} reports the performance measures for every annotator in labeling phase 2.

Compared to Table \ref{paper1_prec_rec_f1_annotator_1} of phase 1, we see that the overall F1 score on sentence level increases by more than 13 percentage points from 63\% to 76.2\% which is solely driven by a 14 percentage points increase in the average recall, as the precision remains similar. This is an indication that the session with the annotators between phase 1 and 2 was successful, even if there is still a gap between precision and recall. On document level we still see the higher scores compared to the sentence aggregation level. However, the performance with respect to the first phase slightly decreased and the gap to the performance with sentence aggregation is not as large anymore. The performances between annotators are still comparable as their F1 score varies between 70\% and 80.6\% on sentence level, with the exception of the finance professor who helped me developing the labeling rules (F1 of 88.6\%).

\begin{table}
\begin{flushleft}
\caption{Annotator Performance Metrics in Labeling Phase 2}
 \mytablefont{In labeling phase 2, the gold labels among all common texts are created by one annotator. This table displays macro precision, recall and F1 of the remaining eight annotators (as defined by \cite{sokolova2009systematic}, averaged over topics) as percentage numbers. Additionally, the table displays the number of texts for each annotator. I conduct the analysis on sentence and document level. Annotators are sorted by decreasing F1 score on sentence level. I remove all topics for which there are less than 3 observations available on sentence level. These topics are \textit{Large Scale Project}, \textit{Real Invest},  \textit{Delay},  \textit{Profit Warning}, \textit{SEO} and  \textit{Pharma Good}.}
\label{paper1_prec_rec_f1_annotator_2}
\mytablefont
\begin{tabularx}{\textwidth}{m{2cm} *{9}{Y}}
\toprule
 \textbf{Annotator} & \textbf{A1} & \textbf{A6} & \textbf{A4} & \textbf{A3} & \textbf{A9} & \textbf{A5} & \textbf{A2} & \textbf{A8} & \textbf{Avg.} \\
\midrule
  \multicolumn{10}{c}{\textit{Panel A: Sentence Level}} \\
       \midrule
Precision & 87.2 & 79.1 & 81.6 & 83.7 & 86.4 & 82.5 & 85.1 & 90.5 & 84.5 \\
Recall & 79.1 & 68.3 & 74.5 & 69.7 & 54.5 & 70.7 & 69.6 & 62.3 & 68.6 \\
F1 & 88.6 & 80.6 & 77.1 & 74.9 & 74.8 & 73.6 & 70.1 & 70.0 & 76.2 \\
Num. & \multicolumn{9}{c}{246} \\
 \midrule 
\multicolumn{10}{c}{\textit{Panel B: Document Level}} \\
       \midrule
Precision & 100 & 80.8 & 88.5 & 84.6 & 93.5 & 90.9 & 92.9 & 81.5 & 89.1 \\
Recall & 88.1 & 72.6 & 81.0 & 79.8 & 77.4 & 59.7 & 75.3 & 83.3 & 77.1 \\
F1 & 91.7 & 83.9 & 83.1 & 80.5 & 79.9 & 79.9 & 79.8 & 78.5 & 82.2 \\
Num. & \multicolumn{9}{c}{20} \\
 \bottomrule
\end{tabularx}
\end{flushleft}
\end{table}

\begin{table}
\begin{flushleft}
\caption{Topic Performance Metrics in Labeling Phase 2}
 \mytablefont{This table displays macro precision, recall and F1 (as defined by \cite{sokolova2009systematic}, averaged over annotators) as percentage numbers for all topics. Additionally, the table displays the number of texts for each topic. I conduct the analysis on sentence and document level. In labeling phase 2, the gold labels among all common texts are created by one annotator. Annotators are sorted by decreasing F1 score on sentence level. I remove all topics for which there are less than 3 observations available on sentence level. These topics are \textit{Large Scale Project}, \textit{Real Invest},  \textit{Delay},  \textit{Profit Warning}, \textit{SEO} and  \textit{Pharma Good}.}
\label{paper1_prec_rec_f1_category_2}
\mytablefont
\begin{tabularx}{\textwidth}{m{3.5cm} *{8}{Y}}
\toprule
 & \multicolumn{4}{c}{\textit{Sentence Level}} & \multicolumn{4}{c}{\textit{Document Level}} \\
 & Num. & Prec. & Recall & F1 & Num. & Prec. & Recall & F1 \\
\midrule
Squeeze Out & 3 & 95.0 & 100 & 96.9 & 1 & 93.8 & 100 & 95.8 \\
Dividend & 5 & 100 & 85.0 & 90.6 & 2 & 100 & 81.2 & 87.5 \\
Bankruptcy Proceedings & 3 & 80.4 & 87.5 & 86.7 & 1 & 78.6 & 87.5 & 85.7 \\
Law & 4 & 95.0 & 75.0 & 83.3 & 2 & 100 & 50.0 & 66.7 \\
Split & 3 & 100 & 62.5 & 82.9 & 1 & 100 & 87.5 & 100 \\
Buyback & 6 & 87.8 & 81.2 & 82.5 & 1 & 100 & 100 & 100 \\
Earnings & 51 & 96.3 & 72.8 & 82.2 & 7 & 100 & 94.6 & 97.0 \\
Bankruptcy Filing & 5 & 78.7 & 65.0 & 81.6 & 1 & 91.7 & 75.0 & 94.4 \\
Debt & 3 & 71.2 & 100 & 81.2 & 1 & 62.5 & 100 & 72.9 \\
Delisting & 4 & 100 & 65.6 & 78.6 & 2 & 100 & 50.0 & 66.7 \\
M \& A & 19 & 71.5 & 61.2 & 63.6 & 3 & 87.5 & 87.5 & 86.1 \\
Management & 6 & 100 & 31.2 & 57.8 & 2 & 100 & 50.0 & 77.8 \\
Restructuring & 7 & 59.9 & 48.2 & 48.2 & 2 & 68.8 & 75.0 & 69.2 \\
Guidance & 4 & 51.6 & 25.0 & 38.4 & 3 & 68.8 & 41.7 & 52.0 \\
 \midrule 
 Avg. & 9 & 84.5 & 68.6 & 76.2 & 2 & 89.1 & 77.1 & 82.2 \\
\bottomrule
\end{tabularx}
\end{flushleft}
\end{table}

Looking at Table \ref{paper1_prec_rec_f1_category_2} which shows the performance per topic in labeling phase 2, we see that there are nine topics for which the annotators reach a F1 score above 80\% on sentence level aggregation. In phase 1, there were only 3 topics that fulfilled that property, even though more topics were considered. Additionally, the scores of the most problematic topics from phase 1 are better in phase 2, as the scores of \textit{Bankruptcy Proceedings}, \textit{Debt} and \textit{Bankruptcy Filing} increase by 35, 36.3 and 42.1 percentage points, respectively. This again highlights the impact of the review session with the annotators between phase 1 and 2.

However, it has to be noted that these improvements have to be considered with caution. As the number of observations is rather small, the results for some topics in phase 2 might not be stable and may be driven by single announcements. For example, the annotators' performance for the topics \textit{Management}, \textit{Restructuring} and \textit{Guidance} decreases with respect to phase 1, even though there is no reason to assume that the annotators should systematically perform worse on these topics. Additionally, the effect of increasing average recall and F1 scores could possibly be biased by the missing topics, as these are mostly topics that perform below average in phase 1. Nevertheless, as the performance for most of the topics increases strongly, there is indication for an improved overall annotator performance in phase 2.
\newpage
\subsubsection{Inter-Annotator Agreement} 

\begin{wraptable}{r}{0.48\textwidth}
\caption{Interpretation $\kappa$ Statistic}
 \mytablefont{This table presents the interpretation of the possible values for the $\kappa$ statistic from \cite{landis1977application}.}
\label{paper1_tab_kappa_interpretation}
\mytablefont
\begin{tabular}{m{2cm} m{4.4cm}}
\toprule
\textbf{$\kappa$} & \textbf{Agreement}\\
\midrule
<0 & Less than chance agreement\\
0.01-0.20 & Slight agreement\\
0.21-0.40 & Fair agreement\\
0.41-0.60 & Moderate agreement\\
0.61-0.80 & Substantial agreement\\
0.81-0.99 & Almost perfect agreement\\
\bottomrule
\end{tabular}
\end{wraptable}
Finally, I test the quality of the database by calculating the inter-annotator agreement between all annotators. In the prior sections, I assume that the gold labels are the ground truth of the labeling process. However, this does not have to be true, since these labels are created by humans which error prone, even if the labeling was done with great caution. A common measure for inter-annotator agreement is the $\kappa$ statistic, which is defined as
\begin{equation}
\label{paper1_eqn_kappa}
\kappa=\frac{p_a-p_e}{1-p_e}
\end{equation}
where $p_a$ denotes the observed rate of agreement between two annotators and $p_e$ is the expected rate of agreement if two annotators would make their assignments randomly.
Equation \ref{paper1_eqn_kappa} normalizes the degree of agreement actually attained above chance by the maximum possible attainable degree of agreement over and above predicted by chance. This score is applicable for binary or nominal data. As I have multi-label data, I treat every topic as a single binary data set, hence I compute the $\kappa$ statistic for every single topic. The $\kappa$ statistics that are most often used in the literature are Fleiss' $\kappa$ \citep{fleiss1971measuring} and Cohen's $\kappa$ \citep{cohen1960coefficient}, with the difference that Fleiss' $\kappa$ allows for more than two annotators. As nine annotators are engaged in this study, Fleiss' $\kappa$ is more suitable. Table \ref{paper1_tab_kappa_interpretation} from  \cite{landis1977application} provides a guideline for how to interpret values for Fleiss' $\kappa$. Table \ref{paper1_tab_fleiss} shows the Fleiss' $\kappa$ statistic for all topics, aggregation levels and labeling phases. 

\begin{table}
\begin{flushleft}
\caption{Fleiss' Kappa}
 \mytablefont{This table measures inter-annotator agreement using Fleiss' Kappa (as percentage numbers) for every topic. I conduct the analysis on sentence and document level for both annotation phases. Phase 1 is conducted with eight annotators since two annotators worked together. Phase 2 is conducted with nine annotators. Topics are sorted by decreasing Fleiss' kappa score on sentence level in phase 1. The topics \textit{Large Scale Project}, \textit{Real Invest},  \textit{Delay},  \textit{Profit Warning}, \textit{SEO} and  \textit{Pharma Good} are removed in phase 2 due to bad coverage.}
\label{paper1_tab_fleiss}
\mytablefont
\begin{tabularx}{\textwidth}{m{3.5cm} *{4}{Y}}
\toprule
 & \multicolumn{2}{c}{\textit{Phase 1}} & \multicolumn{2}{c}{\textit{Phase 2}} \\
 & Sentence & Document & Sentence & Document \\
\midrule
Large Scale Project & 94.0 & 100.0 & - & - \\
Dividend & 82.0 & 96.3 & 88.3 & 83.6 \\
Management & 79.9 & 88.5 & 54.3 & 57.6 \\
Squeeze Out & 77.7 & 88.2 & 93.0 & 89.4 \\
Split & 76.2 & 100.0 & 79.0 & 86.9 \\
Real Invest & 67.8 & 87.3 & - & - \\
Earnings & 66.6 & 83.4 & 74.0 & 93.1 \\
Pharma Good & 65.1 & 86.5 & - & - \\
Delisting & 63.8 & 83.7 & 86.9 & 89.4 \\
Delay & 62.0 & 87.6 & - & - \\
Guidance & 61.6 & 63.6 & 23.1 & 25.9 \\
Law & 60.6 & 87.2 & 89.9 & 89.4 \\
M \& A & 59.6 & 67.8 & 55.6 & 76.7 \\
SEO & 53.3 & 68.1 & - & - \\
Buyback & 51.2 & 68.6 & 79.8 & 100.0 \\
Profit Warning & 48.1 & 58.3 & - & - \\
Bankruptcy Filing & 47.7 & 64.1 & 57.6 & 64.0 \\
Debt & 47.6 & 79.1 & 80.3 & 65.9 \\
Bankruptcy Proceedings & 46.4 & 65.7 & 66.9 & 68.5 \\
Restructuring & 45.8 & 55.2 & 39.1 & 54.2 \\
 \midrule 
 Avg. & 62.9 & 79.0 & 69.1 & 74.6 \\
\bottomrule
\end{tabularx}
\end{flushleft}
\end{table}

First of all, we see that the average agreement for all labeling phases and aggregation levels is substantial, according to \cite{landis1977application}. The general pattern that we have identified for the annotators' performance also repeats for the inter-annotator agreement: The average agreement on sentence level increases by about 6.2 percentage points in phase 2, while the agreement on document level decreases by about 4.4 percentage points. Looking at the single topics, we see that similar topics perform badly as in the annotator performance study. Again, we see a clear increase in agreement for these topics in phase 2.
\newpage
\subsection{Final Corpus}
The final corpus of the ad-hoc multi-label database consists of 31,771 sentences from 3,044 announcements. Table \ref{paper1_tab_sum_stat_database} computes basic summary statistics for the final corpus for all aggregation levels, such the number of texts per announcements, number of labels per text and the number of labels per topic.

\begin{table}
\begin{flushleft}
\caption{Descriptive Statistics Ad-Hoc Multi-Label Database}
 \mytablefont{This table presents the summary statistics of the ad-hoc multi-label database on sentence and document level. I compute selected percentiles as well as the mean, standard deviation and number of observations.}
\label{paper1_tab_sum_stat_database}
\mytablefont
\begin{tabularx}{\textwidth}{c *{8}{Y}}
\toprule
 & Count & Mean & Std & Min & 25\% & 50\% & 75\% & Max \\
\midrule
   \multicolumn{9}{c}{\textbf{Panel A: Sentence Level}} \\
       \midrule
Num. Sentences Per Announcement & 3,044 & 10.4 & 8.3 & 1 & 5 & 9 & 13 & 112 \\
Num. Labels Per Sentence & 31,771 & 0.5 & 0.6 & 0 & 0 & 1 & 1 & 4 \\
Num. Labels Per Topic & 20 & 864.1 & 828.9 & 287 & 536 & 648 & 764 & 4,218 \\
\midrule
   \multicolumn{9}{c}{\textbf{Panel B: Document Level}} \\
       \midrule
Num. Documents Per Announcement & 3,044 & 1.0 & 0.0 & 1 & 1 & 1 & 1 & 1 \\
Num. Labels Per Document & 3,044 & 1.7 & 1.0 & 0 & 1 & 1 & 2 & 7 \\
Num. Labels Per Topic & 20 & 256.6 & 157.9 & 112 & 171 & 210 & 255 & 748 \\
\bottomrule
\end{tabularx}
\end{flushleft}
\end{table}

Looking at the number of sentences per announcement, we see that the database consists of announcements with 10.4 sentences on average. The 25. and 75. percentile are 5 and 13, respectively. There is one outlier in the corpus with 112 sentences.
On average, there is one label every two sentences. However, it is even possible that one sentence has four labels. The average number of labels is higher on document level with a value of 1.7, with a maximum of seven labels per document.
Looking at the topics on sentence level, we see that there are on average 864.1 labeled sentences per topic within the whole database. However, this value is very volatile, ranging from 287 to 4,218 labeled sentences per document. This pattern repeats also on document level.

\begin{figure}[t]
\caption{Label Distribution Across Topics \& Top 10 Most Co-Occurring Topics}
\subfloat[\centering \mytablefont{Label Distribution Across Topics (Sentence Level)}]
{\includegraphics[width=\linewidth]{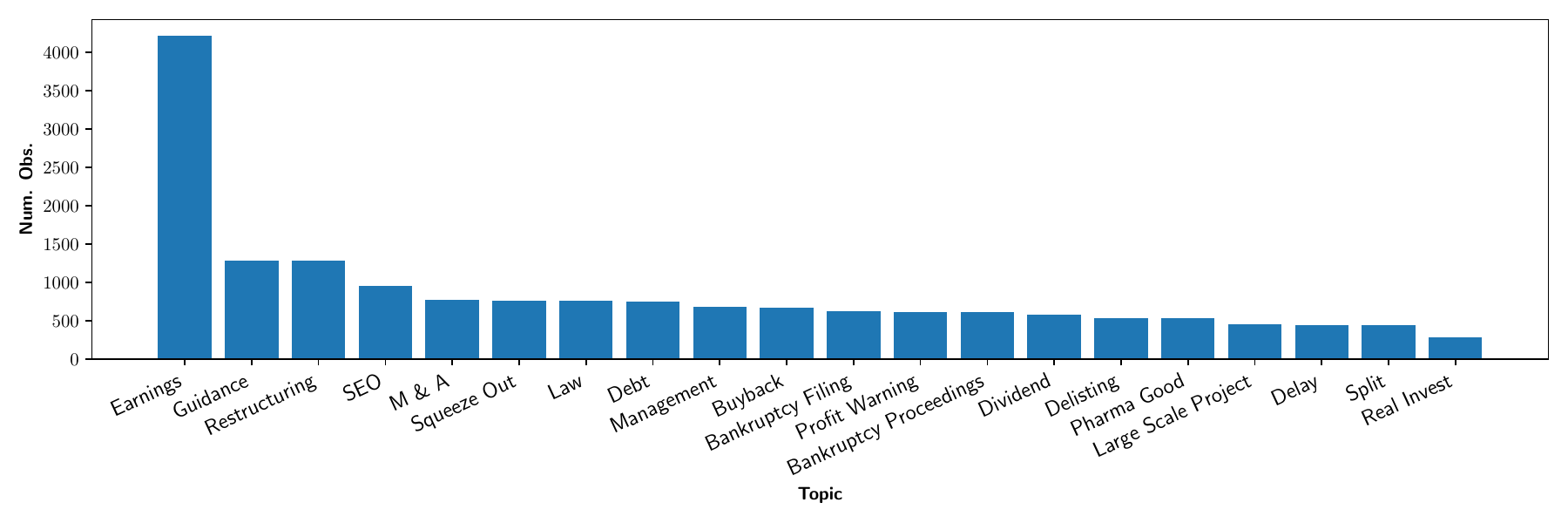}}\par
\subfloat[\centering \mytablefont{Top 10 Most Co-Occurring Topics (Document Level)}]
{\includegraphics[width=\linewidth]{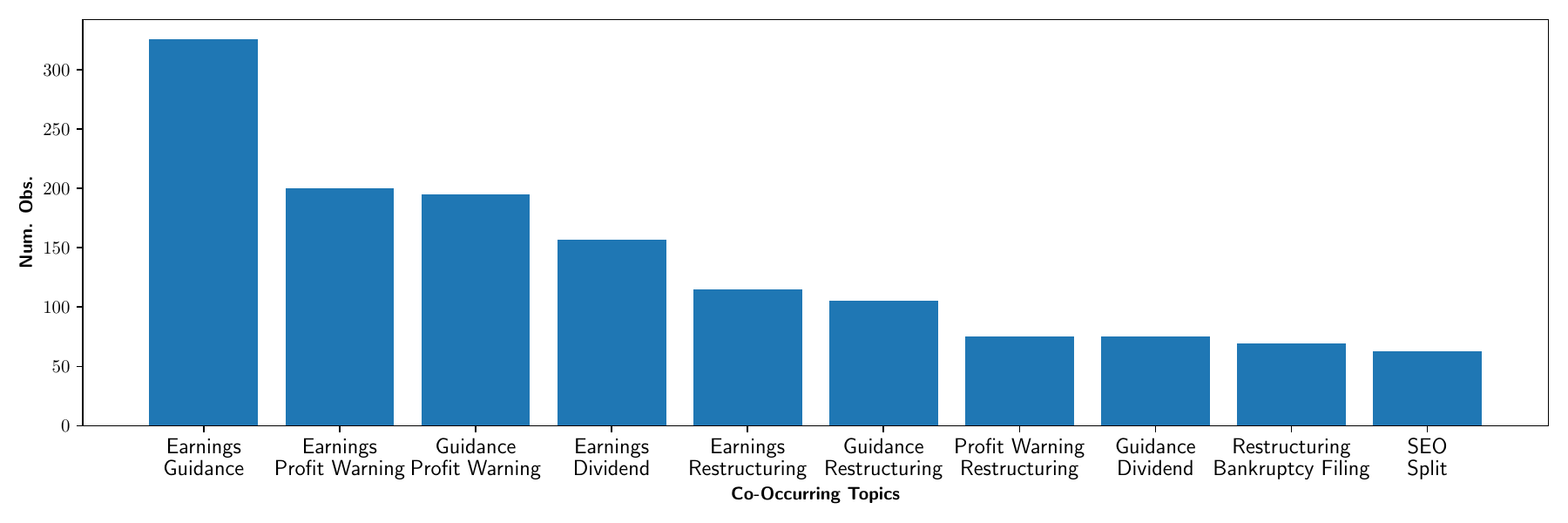}}\par
\label{paper1_fig_topic_hist_cooccurrence_gold}
\mytablefont{Part (a) of this figure is a histogram of the topics in the final ad-hoc multi-label database on sentence level. Part (b) illustrates the ten most frequent topic pairs among all documents within the ad-hoc multi-label database.}
\end{figure}

Figure \ref{paper1_fig_topic_hist_cooccurrence_gold} (a) illustrates the label distribution across topics of the database on sentence level. It is noticeable that the \textit{Earnings} topic with 4,218 observations has more than three times more labeled sentences as the next largest topic, which is the \textit{Guidance} topic with 1,285 observations. Other large topics are \textit{Restructuring} (1282) and \textit{SEO} (959). The smallest topics are \textit{Delay} (446), \textit{Split} (442) and \textit{Real Invest} (287). This implies that even after conducting efforts to balance the database, it is still unbalanced to a certain degree. Figure \ref{paper1_fig_topic_hist_cooccurrence_gold} (b) presents the ten most frequent co-occurring topic pairs. Here we see that many topics co-occur with frequent topics like \textit{Earnings}, \textit{Guidance} and \textit{Restructuring}. This partly explains the unbalanced distribution of the topics, as these frequent topics often co-occur when the BM25 pre-label of an announcement was initially referring to another topic. From the 3,044 documents in the database, 61 have no label and 1,346 have more than one label, which provides evidence for the validity of the multi-label approach.

As a final step, I translate all sentences of the German ad-hoc multi-label database to English, using the translation tool \textit{DeepL}\footnote{https://www.deepl.com/translator}. This allows to train a model that is able to classify English financial announcements, as most of the financial text data is given in English.

\section{Ad-Hoc Classification}
\label{paper1_adHocClassification}    
\subsection{Model}
The model that I fit to the ad-hoc multi-label database is the BERT model (Bidirectional Encoder Representations from Transformers) from \cite{bert}. BERT is a deep learning model for natural language processing tasks. It works by pre-training stacked encoder layers of a transformer network \citep{transformer} on a large corpus of text data. This pre-training allows BERT to learn rich and diverse semantic information about the language.

BERT is a sequence-to-sequence model, which transforms a sequence of words or tokens into a sequence of contextualized word embeddings. However, to allow for various downstream-tasks like text classification, the authors add a special token at the beginning of each input sequence, called the [CLS] token, which stands for \textit{classification}. The respective final embedding of this token is used as a representation of the entire sentence and can be used for further tasks during fine-tuning. BERT models use a specific tokenizer which splits words in common sub-words. In that way, the authors circumvent the possibility of out-of-vocabulary words while maintaining a reasonable dictionary size of 30,000 tokens.

Since the ad-hoc multi-label database consists of German data, I use the German version of the BERT-base cased model\footnote{https://huggingface.co/bert-base-german-cased}. This version of BERT is composed of 12 transformer encoder layers with 768 hidden units and 12 self-attention heads. The model is pre-trained on a corpus with more than 12 GB of textual data from German Wikipedia, the OpenLegalData dump and news articles using the masked language model pre-training objective together with the next sentence prediction task. I fine-tune the pre-trained version of BERT on the ad-hoc multi-label database. To that end, I pass the [CLS] embedding to a two-layer feed-forward neural network with hyperbolic tangent and sigmoid activation functions, respectively. The final layer has 20 output neurons and is, together with the sigmoid activation function, suited for the multi-label nature of the database. During inference, the model predicts a specific topic if the respective output is larger than a threshold of 0.6. I test several thresholds, but 0.6 yields the best results.
\subsection{Benchmarks}
The first benchmark model I use in this study is a dense two-layer feed-forward neural network (\textit{NN}) with trainable, 64-dimensional and randomly initialized word embeddings, similar to the fasttext model of \cite{joulin2016bag}. The sequence of word embeddings coming from the embedding layer is pooled using max-pooling and normalized with batch normalization \citep{ioffe2015batch}. The output is then passed through two dense layers with respective output sizes of 64 and 20 and with rectified linear unit and sigmoid activation functions, respectively.

The second benchmark is a two-layer recurrent neural network with gated recurrent unit (\textit{GRU}) layers as defined in \cite{GRU}. Similar to the first benchmark, the initial word embeddings are randomly initialized but with 256-dimensional embeddings. Both GRU-layers keep the embedding size on 256 and use hyperbolic tangent activation functions. All recurrent neural networks are sequence to sequence models, similar to BERT. This means that the output of the GRU-layers is a sequence of word embeddings. However, I only use the embedding of the last token of GRU-output as this embedding contains all the information of previous words. This embedding is passed to a dense layer with output size 20 and sigmoid activation function.

The last benchmark is a two-layer bidirectional GRU (\textit{Bi-GRU}) model with 300-dimensional, pre-trained fasttext word vectors\footnote{https://fasttext.cc/docs/en/crawl-vectors.html}, inspired by the ELMo model of \cite{ELMO}. The bidirectional GRU layers consist of two standard GRU layers with 80-dimensional outputs and hyperbolic tangent activation functions. The difference between both GRU layers is that the input tokens are fed to the one layer in forward direction and to the other in backward direction. The token embeddings of both GRU layers are then concatenated to one embedding of dimension 160 for every token. The Bi-GRU output is pooled twice: Once with a max-pooling and once with an average pooling. Both pooled embeddings are concatenated to a 320-dimensional array, which is finally passed to a dense layer with output size 20 and sigmoid activation function. For the first two benchmarks, I tokenize text into words and I use the 20,000 most frequent words as a dictionary. I treat missing words as if they would not be present.

The main features of the BERT model are that it is pre-trained on a large amount of data and thus is suitable for transfer learning, it considers both the context to the left and right of a word, it uses sub-word tokenization which prevents out-of-vocabulary words and it can be trained efficiently through parallelization. In contrast, the first baseline, the neural network model, has none of these features. The standard GRU model adds a context dependency to the word embeddings but only unidirectional.  Like BERT, the Bi-GRU model has bidirectional contextualized word embeddings and it has pre-trained initial word embeddings through the given fasttext word vectors. However, the Bi-GRU model cannot be pre-trained on a comparably large data set as it is not parallelizable due to its recurrent nature. Also the sub-word tokenization is not applied. Thus, the benchmark models gradually add specific features of the BERT model. In that way, I can observe which feature adds any benefit to the model's performance on the specific multi-label classification task.
\subsection{Training}
I train the BERT model and all benchmarks on both sentence and document level data. This methodology enables an exploration into whether training on sentence-level data yields superior topic models compared to the conventional approach of utilizing document-level data. I use the adaptive moment estimation method of \cite{kingma2014adam}, called Adam, to fine-tune the BERT-base model with a batch size of six for four epochs. I vary the learning rate and the first exponential decay rate $\beta_1$ during training using a triangular cyclical learning rate policy as proposed by \cite{smith2017cyclical}, as the authors get an increased model performance with fewer training iterations compared to standard training approaches. To find valuable lower and upper bounds for the learning rate, the authors propose a learning rate range test which trains the model for several epochs while letting the learning rate increase linearly between low and high learning rate values. That test yields a maximum learning rate of $2\mathrm{e}-5$ and a minimum learning rate of $2\mathrm{e}-6$. A later paper of the same authors further improves the cyclical learning rate schedule by proposing the \textit{1cycle policy} \citep{smith2019super}. This policy proposes to only use one cycle during the whole training with a cycle length that is equal to the total number of iterations during training. In my case, that means that the learning rate, starting from $2\mathrm{e}-6$, increases linearly up to the upper bound after two epochs. Subsequently, it decreases linearly down to the lower bound again after four epochs. The same holds for the first exponential decay rate $\beta_1$, only in reverse order (decreases first, increases last).

I train the model eight times with different shuffled training batches and I report all results as averages of the performance measures of these eight models. In that way, I reduce the likelihood of ending with good results by chance. 
For all other models and aggregation levels, I repeat this approach, as the optimal hyperparameters choices differ between models and data sets. For all models, I use the binary cross-entropy loss function.
\subsection{Classification Results}
\label{paper_1_classification_results}
Table \ref{paper1_tab_model_performance} presents the out-of-sample performance of all models, evaluated on document level. I train every model twice, once with inputs on sentence level and once with inputs on document level. For inputs on sentence level, I aggregate the topic predictions to end up with topic predictions on document level. The main overall performance measure I focus on is the macro F1 score \citep{sokolova2009systematic} since this score is the average of the individual F1 scores of all topics. This means that every topic gets the same weight and is treated as equally important. Given that the primary objective is to develop a model capable of accurately predicting each topic, the macro F1 score serves as a valid metric for comparing the performance of the models in alignment with this goal. However, I additionally report a second overall performance measure,~i.e. the micro F1 score. This score is also an average of the F1 scores of all topics, but weighted with respect to their frequency in the sample. Thus, the micro F1 score indicates how well the model is able to predict especially frequent classes. The inclusion of this measure in the analysis serves to identify potential difficulties the model may encounter in learning patterns for less frequent and minor topics. Such challenges are evidenced by a substantial disparity between the macro and micro F1 scores.

\begin{table}
\begin{flushleft}
\caption{Ad-Hoc Topic Model Performance}
 \mytablefont{This table compares the multi-label classification performance of the cased version of the German BERT base model with an ELMo model with GRU layers and pre-trained, fasttext word embeddings \citep{ELMO}, a standard GRU model \citep{GRU} and a two-layer feed-forward neural network. The latter two models use randomly initialized word embeddings. Every model is trained for eight different seeds. I report the mean F1 score among seeds for every model and topic together with the standard deviation. In addition, I report the micro and macro F1 scores \citep{sokolova2009systematic} as overall performance measures and the number of labels per topic. I evaluate the performance on document level. I report the performance of models that are trained on sentence \& document level. Numbers are given in percent.}
\label{paper1_tab_model_performance}
\mytablefont
\begin{tabularx}{\textwidth}{c *{10}{Y}}
\toprule
 & \multicolumn{5}{c}{\textit{Sentence Level}} & \multicolumn{5}{c}{\textit{Document Level}} \\
 & Num. & BERT & BiGRU & GRU & NN & Num. & BERT & BiGRU & GRU & NN \\
\midrule
Avg. (Macro) & 3,480 & \shortstack[c]{85.3 \\ (0.5)} & \shortstack[c]{78.2 \\ (0.6)} & \shortstack[c]{37.2 \\ (6.2)} & \shortstack[c]{64.4 \\ (1.2)} & 1,029 & \shortstack[c]{78.6 \\ (0.9)} & \shortstack[c]{72.1 \\ (1.3)} & \shortstack[c]{11.6 \\ (1.1)} & \shortstack[c]{19.9 \\ (1.5)} \\[0.25cm]
Avg. (Micro) & 3,480 & \shortstack[c]{85.0 \\ (0.6)} & \shortstack[c]{79.2 \\ (0.4)} & \shortstack[c]{53.3 \\ (4.7)} & \shortstack[c]{70.2 \\ (0.9)} & 1,029 & \shortstack[c]{78.6 \\ (0.8)} & \shortstack[c]{73.5 \\ (0.8)} & \shortstack[c]{31.4 \\ (0.8)} & \shortstack[c]{37.1 \\ (1.8)} \\ 
 \midrule
Squeeze Out & 152 & \shortstack[c]{96.2 \\ (0.6)} & \shortstack[c]{94.0 \\ (0.5)} & \shortstack[c]{91.7 \\ (1.1)} & \shortstack[c]{89.9 \\ (1.1)} & 45 & \shortstack[c]{93.3 \\ (0)} & \shortstack[c]{93.0 \\ (1.1)} & \shortstack[c]{82.5 \\ (4.3)} & \shortstack[c]{93.1 \\ (0.5)} \\[0.25cm]
Delay & 77 & \shortstack[c]{93.8 \\ (1.2)} & \shortstack[c]{87.3 \\ (2.0)} & \shortstack[c]{20.2 \\ (23.6)} & \shortstack[c]{75.3 \\ (6.5)} & 33 & \shortstack[c]{83.5 \\ (2.9)} & \shortstack[c]{76.8 \\ (3.6)} & \shortstack[c]{0 \\ (0)} & \shortstack[c]{0 \\ (0)} \\[0.25cm]
Delisting & 106 & \shortstack[c]{92.4 \\ (1.6)} & \shortstack[c]{89.4 \\ (2.6)} & \shortstack[c]{75.2 \\ (8.0)} & \shortstack[c]{87.1 \\ (1.5)} & 35 & \shortstack[c]{86.3 \\ (2.1)} & \shortstack[c]{86.1 \\ (4.2)} & \shortstack[c]{3.8 \\ (6.8)} & \shortstack[c]{18.8 \\ (20.5)} \\[0.25cm]
Management & 133 & \shortstack[c]{92.2 \\ (1.1)} & \shortstack[c]{87.2 \\ (1.0)} & \shortstack[c]{34.5 \\ (29.5)} & \shortstack[c]{84.7 \\ (4.0)} & 52 & \shortstack[c]{86.5 \\ (1.6)} & \shortstack[c]{82.9 \\ (2.8)} & \shortstack[c]{10.8 \\ (18.2)} & \shortstack[c]{49.3 \\ (20.1)} \\[0.25cm]
Law & 126 & \shortstack[c]{91.2 \\ (1.6)} & \shortstack[c]{86.0 \\ (4.0)} & \shortstack[c]{0 \\ (0)} & \shortstack[c]{55.0 \\ (12.9)} & 38 & \shortstack[c]{84.6 \\ (3.3)} & \shortstack[c]{82.7 \\ (2.8)} & \shortstack[c]{0 \\ (0)} & \shortstack[c]{0 \\ (0)} \\[0.25cm]
Dividend & 115 & \shortstack[c]{91.1 \\ (1.3)} & \shortstack[c]{91.6 \\ (1.5)} & \shortstack[c]{90.7 \\ (1.6)} & \shortstack[c]{93.6 \\ (0.6)} & 50 & \shortstack[c]{82.9 \\ (2.5)} & \shortstack[c]{86.4 \\ (3.3)} & \shortstack[c]{0 \\ (0)} & \shortstack[c]{7.0 \\ (14.1)} \\[0.25cm]
Debt & 161 & \shortstack[c]{90.0 \\ (2.6)} & \shortstack[c]{84.8 \\ (3.2)} & \shortstack[c]{33.6 \\ (20.7)} & \shortstack[c]{65.3 \\ (6.4)} & 39 & \shortstack[c]{84.0 \\ (2.3)} & \shortstack[c]{77.0 \\ (2.6)} & \shortstack[c]{0 \\ (0)} & \shortstack[c]{0 \\ (0)} \\[0.25cm]
Earnings & 914 & \shortstack[c]{89.4 \\ (0.8)} & \shortstack[c]{87.0 \\ (1.3)} & \shortstack[c]{85.9 \\ (1.4)} & \shortstack[c]{84.1 \\ (1.4)} & 150 & \shortstack[c]{87.9 \\ (0.8)} & \shortstack[c]{85.1 \\ (1.1)} & \shortstack[c]{81.8 \\ (1.3)} & \shortstack[c]{78.7 \\ (2.1)} \\[0.25cm]
Buyback & 132 & \shortstack[c]{88.2 \\ (2.7)} & \shortstack[c]{86.1 \\ (4.5)} & \shortstack[c]{31.6 \\ (31.8)} & \shortstack[c]{81.4 \\ (4.0)} & 29 & \shortstack[c]{88.7 \\ (1.8)} & \shortstack[c]{81.7 \\ (3.6)} & \shortstack[c]{0 \\ (0)} & \shortstack[c]{0.8 \\ (2.4)} \\[0.25cm]
Split & 86 & \shortstack[c]{87.6 \\ (2.1)} & \shortstack[c]{84.8 \\ (3.4)} & \shortstack[c]{42.6 \\ (27.1)} & \shortstack[c]{77.9 \\ (2.2)} & 29 & \shortstack[c]{81.0 \\ (2.2)} & \shortstack[c]{78.9 \\ (3.7)} & \shortstack[c]{0 \\ (0)} & \shortstack[c]{2.3 \\ (6.4)} \\[0.25cm]
Pharma Good & 97 & \shortstack[c]{87.5 \\ (3.4)} & \shortstack[c]{87.5 \\ (4.0)} & \shortstack[c]{13.7 \\ (22.7)} & \shortstack[c]{79.8 \\ (12.5)} & 22 & \shortstack[c]{85.7 \\ (1.7)} & \shortstack[c]{86.6 \\ (2.9)} & \shortstack[c]{0 \\ (0)} & \shortstack[c]{0 \\ (0)} \\[0.25cm]
Bankruptcy Filing & 118 & \shortstack[c]{87.2 \\ (1.6)} & \shortstack[c]{80.8 \\ (1.7)} & \shortstack[c]{66.1 \\ (1.2)} & \shortstack[c]{74.2 \\ (1.9)} & 44 & \shortstack[c]{79.4 \\ (2.0)} & \shortstack[c]{73.3 \\ (6.2)} & \shortstack[c]{1.1 \\ (2.0)} & \shortstack[c]{42.1 \\ (17.2)} \\[0.25cm]
Large Scale Project & 93 & \shortstack[c]{86.8 \\ (1.7)} & \shortstack[c]{81.7 \\ (3.6)} & \shortstack[c]{0 \\ (0)} & \shortstack[c]{80.3 \\ (1.1)} & 35 & \shortstack[c]{87.6 \\ (1.9)} & \shortstack[c]{78.2 \\ (1.6)} & \shortstack[c]{0 \\ (0)} & \shortstack[c]{30.3 \\ (27.7)} \\[0.25cm]
SEO & 198 & \shortstack[c]{83.6 \\ (3.1)} & \shortstack[c]{81.6 \\ (3.2)} & \shortstack[c]{56.9 \\ (22.8)} & \shortstack[c]{78.3 \\ (1.3)} & 49 & \shortstack[c]{76.1 \\ (4.1)} & \shortstack[c]{77.4 \\ (2.3)} & \shortstack[c]{5.5 \\ (8.7)} & \shortstack[c]{6.8 \\ (17.8)} \\[0.25cm]
Bankruptcy Proceedings & 125 & \shortstack[c]{81.2 \\ (2.2)} & \shortstack[c]{77.7 \\ (2.5)} & \shortstack[c]{47.1 \\ (22.6)} & \shortstack[c]{73.7 \\ (2.1)} & 40 & \shortstack[c]{76.6 \\ (1.8)} & \shortstack[c]{76.7 \\ (3.7)} & \shortstack[c]{1.2 \\ (3.4)} & \shortstack[c]{21.6 \\ (14.7)} \\[0.25cm]
Restructuring & 245 & \shortstack[c]{79.7 \\ (1.4)} & \shortstack[c]{67.9 \\ (2.3)} & \shortstack[c]{0.3 \\ (0.7)} & \shortstack[c]{46.6 \\ (5.4)} & 97 & \shortstack[c]{69.8 \\ (3.0)} & \shortstack[c]{57.3 \\ (3.4)} & \shortstack[c]{0 \\ (0)} & \shortstack[c]{0 \\ (0)} \\[0.25cm]
Guidance & 253 & \shortstack[c]{78.8 \\ (1.5)} & \shortstack[c]{73.5 \\ (1.7)} & \shortstack[c]{54.6 \\ (9.6)} & \shortstack[c]{61.1 \\ (7.0)} & 108 & \shortstack[c]{67.3 \\ (1.8)} & \shortstack[c]{62.9 \\ (1.3)} & \shortstack[c]{44.4 \\ (2.7)} & \shortstack[c]{46.5 \\ (9.2)} \\[0.25cm]
M \& A & 140 & \shortstack[c]{72.1 \\ (3.8)} & \shortstack[c]{56.4 \\ (6.1)} & \shortstack[c]{0 \\ (0)} & \shortstack[c]{0 \\ (0)} & 49 & \shortstack[c]{66.4 \\ (4.0)} & \shortstack[c]{41.9 \\ (8.3)} & \shortstack[c]{0 \\ (0)} & \shortstack[c]{0 \\ (0)} \\[0.25cm]
Real Invest & 72 & \shortstack[c]{70.2 \\ (3.3)} & \shortstack[c]{30.9 \\ (7.6)} & \shortstack[c]{0 \\ (0)} & \shortstack[c]{0 \\ (0)} & 25 & \shortstack[c]{51.6 \\ (5.9)} & \shortstack[c]{11.7 \\ (6.8)} & \shortstack[c]{0 \\ (0)} & \shortstack[c]{0 \\ (0)} \\[0.25cm]
Profit Warning & 137 & \shortstack[c]{67.0 \\ (2.5)} & \shortstack[c]{48.8 \\ (3.1)} & \shortstack[c]{0 \\ (0)} & \shortstack[c]{0 \\ (0)} & 60 & \shortstack[c]{52.0 \\ (5.1)} & \shortstack[c]{45.2 \\ (6.7)} & \shortstack[c]{0 \\ (0)} & \shortstack[c]{0 \\ (0)} \\
\bottomrule
\end{tabularx}
\end{flushleft}
\end{table}

For the models that are trained on sentence level input, Table \ref{paper1_tab_model_performance} reports that the BERT model achieves an average macro F1 score of 85.3\%, which is more than 7 percentage points better than the next best benchmark, the Bi-GRU model with a macro F1 score of 78.2\%. The NN model has a macro F1 score of 64.4\% and the GRU model achieves only 37.2\%. The small deltas between the macro and micro F1 scores for BERT and Bi-GRU (-0.3 \& 1 percentage points, respectively) indicate that these models perform well also for infrequent topics, which is not the case for GRU and NN (16.1 \& 5.8 percentage points, respectively). The performances of BERT, Bi-GRU and NN are relatively stable as the low standard deviations of the F1 scores imply a robustness against shuffling of the input data and random weight initialization. Looking at the F1 scores of the single topics, we see that the BERT model outperforms all other models in almost all topics. Only for the \textit{Dividend} topic, the Bi-GRU performs slightly better (0.5 percentage points). There is a high discrepancy between the F1 scores of the topics, as the BERT model performs best for the \textit{Squeeze Out} topic with 96.2\% and worst for the \textit{Profit Warning} topic with 67\%. However, there are only four more topics for which the BERT model produces a F1 score below 80\%. These topics are \textit{Restructuring} (79.7\%), \textit{Guidance} (78.8\%), \textit{M \& A} (72.1\%) and \textit{Real Invest} (70.2\%). Reasons for that might be bad coverage (\textit{Real Invest}) or ambiguous topic definitions (\textit{Restructuring}, \textit{Profit Warning} \& \textit{Guidance}). Table \ref{paper1_prec_rec_f1_annotator_2} indicates that even human annotators have problems with topics like \textit{Restructuring} and \textit{Guidance}, so it is not surprising that the model has problems, too. For the Bi-GRU model, the pattern that describes which topics work well and which are problematic is similar. However, especially the problematic topics are even worse. This effect is even stronger for GRU and NN since their F1 scores of 0 for these topics indicate that no pattern is learned at all.

Looking at the performances of models trained on document level input, we see that the general structure of the results we found on sentence level also holds on document level. However, there is a large gap in the absolute performances. For instance, the average macro F1 score of the BERT model is almost 7 percentage points lower if it is trained on document level instead of sentence level input. While this gap is similar for the Bi-GRU model, it is even larger for the GRU (25.6 percentage points) and the NN (44.5 percentage points).

In summary, the BERT model outperforms all benchmarks. The only model that comes close is the Bi-GRU model. This indicates that pre-training and the bidirectional context in word embeddings play a key role in the model performance in this multi-label data set. The value of these features gets especially clear if we compare the Bi-GRU model with the standard GRU model, which are structurally similar. We also see that there is a large performance drop if the models are trained on document level data, which implies that, at least for the multi-label topic classification problem, the common approach of labeling whole documents is not optimal.

Table \ref{paper1_tab_model_precRecF1} in the appendix separates the F1 scores of the BERT model into the respective precision and recall scores. This analysis reveals that the BERT model tends to label too frequently. However, this is a direct consequence of the label structure of the ad-hoc multi-label database. See the appendix for further explanations.

Finally, as most of the financial text is given in English, I train English versions of the BERT model and all benchmarks with exactly the same settings as in the German case. The results, depicted in Table \ref{paper1_tab_model_performance_en} of the appendix, reveal that the English version of the BERT topic model performs on par with its German counterpart. 

\section{The Impact of Ad-Hoc Announcement Topics on Stock Prices}
\label{paper1_sec_StockMarketReaction}
In this section, the study examines the stock price responses to ad-hoc announcements in relation to their respective topics. To that end, I begin with the estimation of the abnormal returns at the report dates of the ad-hoc announcements with a one-day event window using the market model:
\begin{equation}
\label{paper1_eqn_marketModel}
r^i_t-r^f_t=\alpha_0+\alpha_1(r^m_t-r^f_t)+\sum_j \alpha_2^{i,j} D^{i,j}_t+\epsilon^i_t,
\end{equation}
where $r^i$ is the return of company $i$, $r^f$ is the risk-free rate (3 months EURIBOR), $r^m$ is the return of the CDAX index (a composite index that contains all stocks traded on the Frankfurt Stock Exchange that are listed in the General or Prime Standard market segments) and $D^{i,j}$ is the event dummy of company $i$ at report date $j$. Accordingly, $\alpha_2^{i,j}$ is the abnormal return of company $i$ and ad-hoc announcement date $j$. The estimation window is one year prior to the report date of the ad-hoc announcement. However, in case that there is not a full year of historical stock returns available for a company, the minimum requirement is 73 days of historical stock returns. That requirement reduces the sample of available ad-hoc announcements, together with the constraint of only using news in German language, from 132,371 to 29,143 observations. I gather the market data from Thomson Reuters Datastream.

\begin{figure}[t]
\caption{Stock Market Reactions on Ad-Hoc Announcement Per Topic}
\subfloat[\centering \mytablefont{Boxplot Abnormal Returns Per Topic}]
{\includegraphics[width=\linewidth]{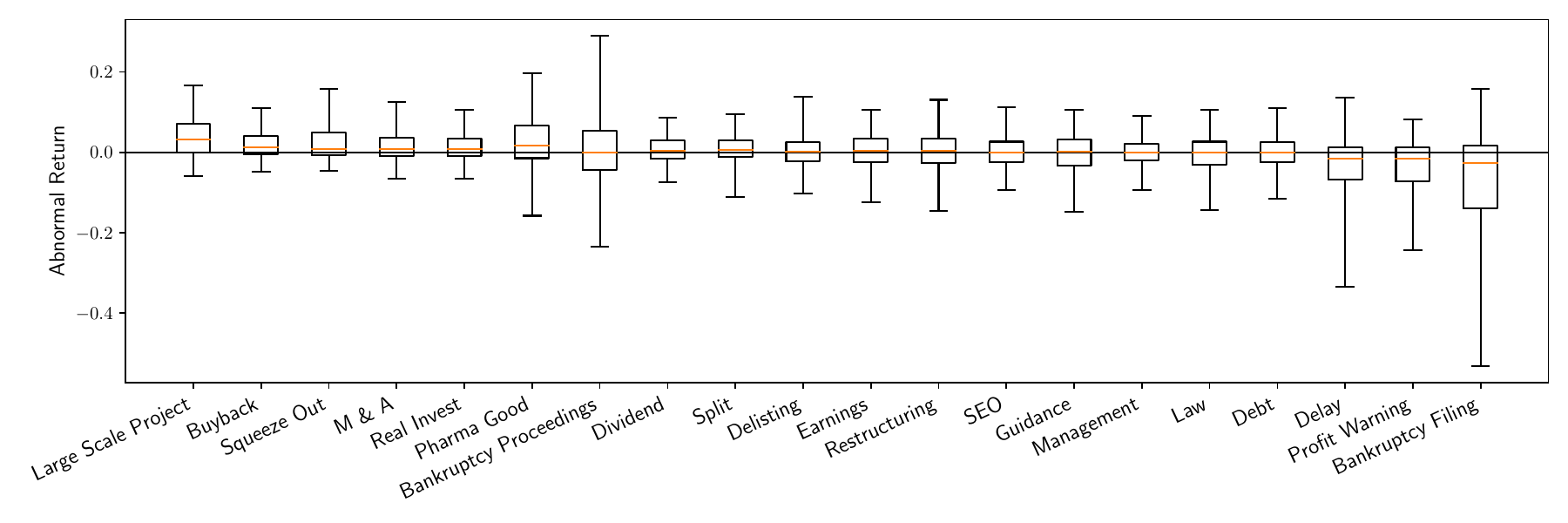}}\par
\subfloat[\centering \mytablefont{Relative Frequency of Significant Positive/Negative Market Reactions Per Topic}]
{\includegraphics[width=\linewidth]{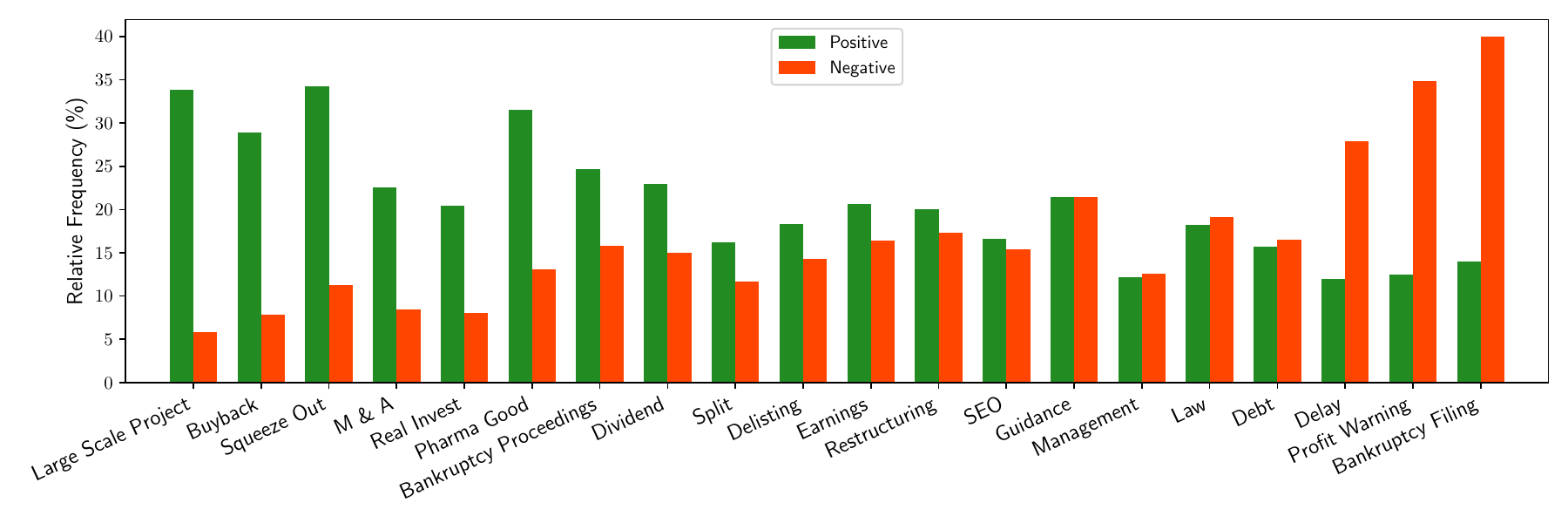}}\par
\label{paper1_fig_stockMarketReactions}
\mytablefont{Part (a) of this figure illustrates the boxplots of the abnormal returns per topic, resulting from the market model regression (\ref{paper1_eqn_marketModel}). The vertical lines represent, from bottom to top, the 5., the 25., the 50. (orange line), the 75. and the 95. percentile. Part (b) plots the relative frequency of statistically significant (with significance level of 10\%) positive and negative abnormal returns on ad-hoc announcements per topic. I leave out the statistically insignificant abnormal returns.}
\end{figure}
Figure \ref{paper1_fig_stockMarketReactions} (a) shows the boxplots of the abnormal returns $\alpha_2^{i,j}$ for each topic. The boxes represent the range between the first and third quartile. The lower and upper ends of the whiskers are the 5. and 95. percentile, respectively. The orange vertical lines represent the median values. These median values indicate that there are substantial differences across topics with respect to abnormal returns. Topics like \textit{Large Scale Project} or \textit{Buyback} clearly produce positive median abnormal returns, with first quartiles being close to zero. This is not surprising: If a company announces a new large-scale project, it can generate a new source of income which can help the company grow its revenue and market share. If a company announces a stock buyback program, the offered repurchase price is generally above the actual stock price so that the actual stockholders have an incentive to sell their shares. In contrast, topics like \textit{Bankruptcy Filing}, \textit{Profit Warning} or \textit{Delay} induce clearly negative median abnormal returns, with third quartiles being close to zero. A delayed mandatory report or a profit warning are indicators that a company is in financial distress, or even worse in case of a bankruptcy filing. If we look at the range between the upper and lower ends of the whiskers, we can see that the volatility of abnormal returns highly depends on the topic. Abnormal returns of ad-hoc announcements containing information about \textit{Buyback}, \textit{Real Invest}, \textit{Dividend} or \textit{Management} are relatively stable compared to announcements that refer to \textit{Pharma Good} or rather negative topics as \textit{Bankruptcy Proceedings}, \textit{Delay}, \textit{Profit Warning} or \textit{Bankruptcy Filing}.  One possible explanation for the high volatility in the negative topics is that the level of surprise of the information varies. If, for example, the market already knows that a company has filed bankrupt, this information should already be incorporated in the stock price of the company, so that one would not expect a significant negative price reaction on another announcement about the bankruptcy filing. Another explanation is that these topics might co-occur with other, more positive topics. For example, some announcements about bankruptcy filings also inform that the creditors' meeting approves an insolvency plan (\textit{Bankruptcy Proceedings} topic), which puts the negative news into perspective for the stockholders. The \textit{Pharma Good} topic is a special case, as this news in most of the cases refers to approvals or denials of new drugs of pharmaceutical companies, where the former generally lead to strong positive and the latter to strong negative market reactions.

Figure \ref{paper1_fig_stockMarketReactions} (a) contains all abnormal returns resulting from regression (\ref{paper1_eqn_marketModel}) regardless of whether they are statistically significant or not. Therefore, in Figure \ref{paper1_fig_stockMarketReactions} (b) I additionally inspect the relative frequency of statistically significant (at the 10\% level) abnormal returns per topic, broken down by positive and negative returns. This figure gives an indication of how likely positive or negative stock market reactions are for the different topics of ad-hoc announcements. We see that topics with clearly more positive than negative statistically significant returns roughly coincide with the positive topics in Figure \ref{paper1_fig_stockMarketReactions} (a). The same holds for topics with negative returns. However, for clearly negative topics like bankruptcy filings, we observe that there is nevertheless a high proportion of significantly positive abnormal returns, which might be explained by positive, co-occurring topics. Additionally, we see that topics like \textit{Earnings}, \textit{Guidance}, \textit{SEO} or \textit{Management} induce a roughly similar amount of significantly positive and negative market reactions. These are topics where the pure information of the presence of the topic is not informative for the direction of abnormal returns. For these cases, further aspects are relevant to determine whether such announcements induce positive or negative market reactions. For example: What are the market expectations of reported earnings results? What are the reasons for a capital increase or a management change? For these topics, looking at co-occurring topics might yield meaningful insights, too. For instance, if an earnings announcement co-occurs with a profit warning, one might expect a negative market reaction.

As a further analysis, I investigate the impact of the topics of ad-hoc announcements on their respective abnormal returns using a standard OLS panel regression with year- and firm-fixed effects and clustered standard errors on time and entity level. For every topic, I define a dummy variable that is one if the topic is present in a given announcement and zero otherwise. Due to the multi-label structure of the data, more than one topic dummy might be active per announcement. Therefore, I run two setups: In the first setup, I regress abnormal returns only on the individual topic dummies. In the second setup, I add pairwise interaction effects between the topics, to test whether topic co-occurrences change the influence of single topics. Since I have 20 different topics, there are 190 possible combinations of pairwise co-occurring topics. However, I require at least 20 announcements in the sample for each topic pair, which reduces the number of topic combinations considered in the regression to 116.

\begin{table}
\begin{flushleft}
\caption{Ad-Hoc Topic Regressions}
 \mytablefont{I regress the abnormal returns on ad-hoc announcements (gathered from equation (\ref{paper1_eqn_marketModel})) on the topics of the announcements. I conduct the estimations using a standard OLS panel regression with year- and firm-fixed effects and clustered standard errors. I run two estimations, one without and one with interaction terms of the topic dummies. In the latter case, I only consider topic pairs that co-occur in at least 20 announcements. I report the \textit{within} version of $R^2$. The topic coefficients and the $R^2$ are given as percentage numbers; the number of observations and regressors are given as absolute numbers. I sort results by the topic coefficients for the scenario without interaction effects. I outsource the results for the interaction effects to Figure \ref{paper1_fig_topicCooccurenceHeatmap}.}
\label{paper1_tab_market_reaction_regression}
\mytablefont
\begin{tabularx}{\textwidth}{c *{5}{Y}}
\toprule
 & \multicolumn{2}{c}{\textit{Without Interaction}} & \multicolumn{2}{c}{\textit{With Interaction}} \\
 & Coefficient & P Value & Coefficient & P Value \\
\midrule
Large Scale Project & 2.70 & 0.00 & 3.27 & 0.00 \\
Bankruptcy Proceedings & 1.95 & 19.08 & -1.84 & 33.17 \\
Pharma Good & 1.22 & 16.70 & 2.18 & 6.00 \\
Buyback & 1.21 & 0.00 & 0.81 & 1.17 \\
Squeeze Out & 1.21 & 3.06 & 1.70 & 1.07 \\
M \& A & 1.00 & 0.03 & 0.51 & 21.96 \\
Real Invest & 0.62 & 1.73 & 0.58 & 12.75 \\
Dividend & 0.35 & 5.57 & -0.35 & 31.65 \\
Split & 0.09 & 84.11 & -0.50 & 61.98 \\
Restructuring & -0.28 & 32.48 & -0.30 & 28.10 \\
Earnings & -0.29 & 19.43 & -0.73 & 0.74 \\
Debt & -0.53 & 8.87 & -1.27 & 2.89 \\
Delisting & -0.73 & 7.61 & -1.91 & 0.15 \\
Guidance & -0.84 & 0.01 & -1.09 & 0.08 \\
SEO & -0.90 & 0.00 & -1.58 & 0.00 \\
Management & -1.19 & 0.00 & -1.54 & 0.00 \\
Law & -1.45 & 0.01 & -2.12 & 0.06 \\
Profit Warning & -4.04 & 0.00 & -2.35 & 0.02 \\
Delay & -4.29 & 0.00 & -5.41 & 0.84 \\
Bankruptcy Filing & -8.75 & 0.00 & -16.02 & 0.00 \\
$R^2$ &  \multicolumn{2}{c}{5.01} &  \multicolumn{2}{c}{6.65}\\
Num. Observations &  \multicolumn{2}{c}{29,143} &  \multicolumn{2}{c}{29,143}\\
Num. Regressors &  \multicolumn{2}{c}{20} &  \multicolumn{2}{c}{136}\\
\bottomrule
\end{tabularx}
\end{flushleft}
\end{table}

Table \ref{paper1_tab_market_reaction_regression} presents the results of the regression. If we first consider the results without interaction effects between topics, we see in total six topics with positive and nine topics with negative coefficients that are statistically significant at least at the 10\% level. The overall pattern of Figure \ref{paper1_fig_stockMarketReactions} is confirmed: Topics like \textit{Large Scale Project}, \textit{Buyback} or \textit{Squeeze Out} have a positive and statistically significant impact on abnormal returns (2.7, 1.21 and 1.21 percentage points, respectively). Note that the topics \textit{Bankruptcy Proceedings} and \textit{Pharma Good} have the second and third highest positive impact on abnormal returns with 1.95 and 1.22 percentage points, respectively. However, these effects are not statistically significant. Accordingly, topics like \textit{Law}, \textit{Profit Warning}, \textit{Delay}, or \textit{Bankruptcy Filing} have a significant negative influence on abnormal returns (-1.45, -4.04, -4.29 and -8.75 percentage points, respectively). Furthermore, we observe that some coefficients of topics with a balanced ratio of significant positive and negative abnormal returns (Figure \ref{paper1_tab_market_reaction_regression} (b)), like \textit{Earnings}, \textit{Restructuring} or \textit{Split}, are insignificant. Economically, this is expectable as these topics, if considered in isolation, are neither positive nor negative and the stock market reaction depends on the context. However, that holds not true for all of the topics where the ratio of positive and negative price changes is balanced. The topics \textit{Debt}, \textit{Delisting}, \textit{Guidance}, \textit{SEO} and \textit{Management} induce negative market reactions, whereas \textit{M \& A}, \textit{Real Invest} and \textit{Dividend} imply positive market reactions.

\begin{figure}[t]
\caption{Heat Map Stock Market Reactions on Ad-Hoc Announcements With Co-Occurring Topics}
{\includegraphics[width=\linewidth]{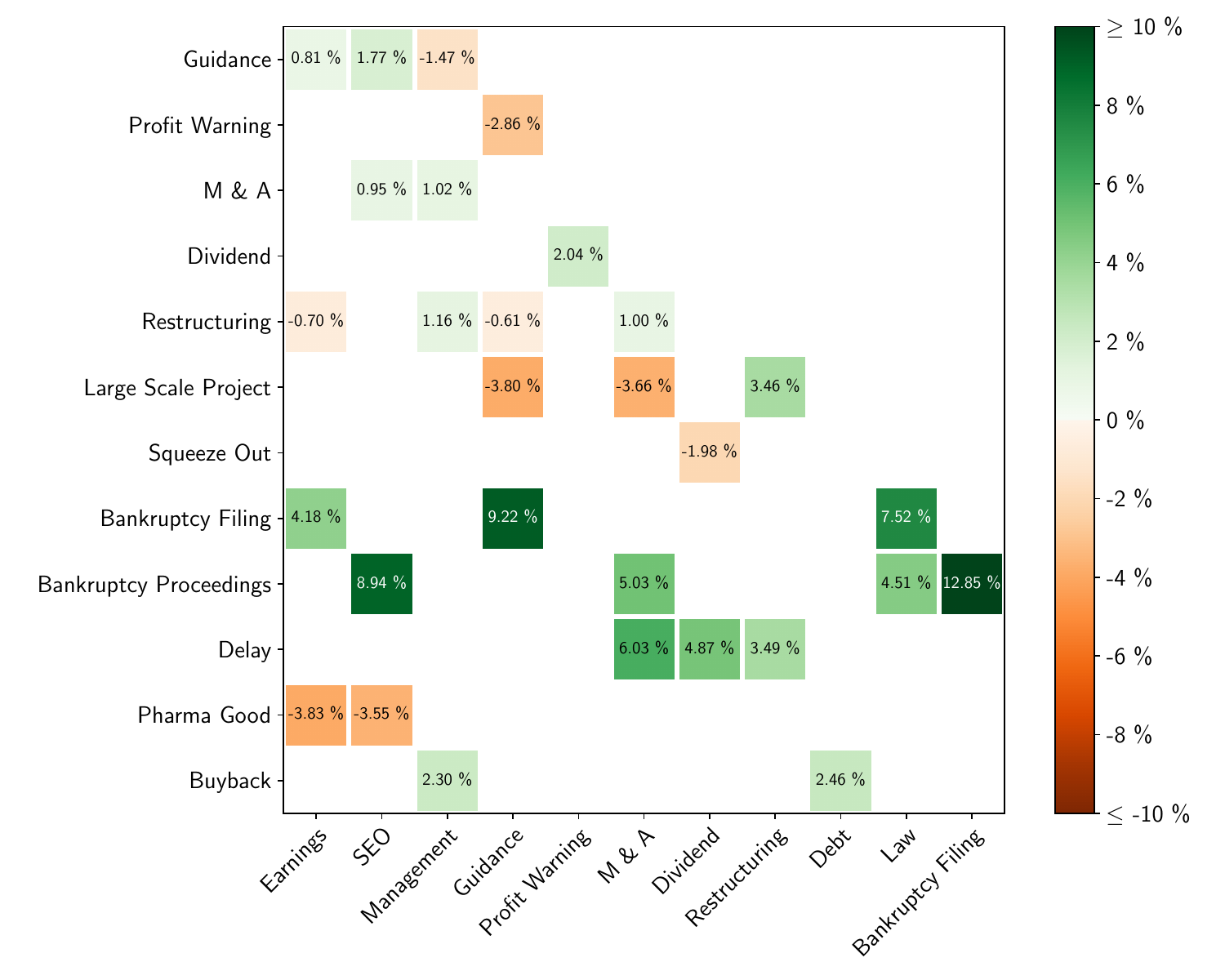}}\par
\mytablefont{This heat map contains all significant interaction effects between the topics. Statistically insignificant effects are left blank. Topics without any significant interaction with any other topic are removed.}
\label{paper1_fig_topicCooccurenceHeatmap}
\end{figure}
In the following, I focus on the results with interaction terms. I outsource the results for the interaction effects to the heat map of Figure \ref{paper1_fig_topicCooccurenceHeatmap} for the sake of readability. It includes all significant interaction effects between the topics. Statistically insignificant effects are left blank. Topics without any significant interaction with any other topic are removed. In the heat map we identify 29 significant interaction effects, which is 25\% of all interaction terms. In Table \ref{paper1_tab_market_reaction_regression}, we see that all of the most positive topics stay positive, even after inclusion of the interaction terms. The same holds for the most negative topics. However, the magnitude of the topic might change, as is the case for instance for \textit{Bankruptcy Filing}. The effect almost doubles from -8.75 percentage points without interactions to -16.02 percentage points with interactions. The reason for that is that I now control for co-occurring topics, which, in case for \textit{Bankruptcy Filings}, formerly positively biased the respective coefficient. Figure \ref{paper1_fig_topicCooccurenceHeatmap} yields evidence for that hypothesis, as the coefficient for the interaction effect between \textit{Bankruptcy Filing} and \textit{Bankruptcy Proceedings} is 12.85\%. Economically, this is reasonable: If an announcement contains both topics, \textit{Bankruptcy Filing} and \textit{Bankruptcy Proceedings}, the information about the filing of bankruptcy is either already known by the market or it is less severe, as for example the financing of the respective company might be secured. This reduces the negative overall effect of the bankruptcy filing significantly. 

However, the more interesting cases are the ambiguous topics where we are not able to economically expect a positive or negative stock market reaction without further information. The previously positive topics \textit{M \& A}, \textit{Real Invest} and \textit{Dividend} are now insignificant. The significantly negative topics stay negative with interaction effects. However, there is now more information to interpret these effects economically. Let us consider for example the topic \textit{SEO}. The coefficient of that topic even decreased from -0.9\% to -1.58\% with the addition of interaction effects. However, looking at co-occurring topics, we see strongly positive (e.g. \textit{Bankruptcy Proceedings}) or negative (e.g. \textit{Pharma Good}) interaction effects. Manually checking ad-hoc announcements that contain these co-occurring topics yields an explanation for these effects: The co-occurring topics give us information about the reason for the capital increase from the \textit{SEO}. If for instance \textit{SEO} co-occurs with \textit{Bankruptcy Proceedings}, the acquired capital from the SEO is very likely used for the restructuring of the company, which is good news for the investor. However, if \textit{SEO} co-occurs with \textit{Pharma Good}, this usually means that the acquired capital is needed to cover additional costs for the development of drugs, which is rather negative news. Furthermore, the formerly insignificant positive coefficient of the \textit{Pharma Good} topic is higher and statistically significant after including the interaction terms. In general, a \textit{Pharma Good} announcement tends to introduce new drugs or their approval, which is positive news for investors of the respective pharmaceutical companies. However, when \textit{Pharma Good} announcements co-occur with \textit{Earnings} or \textit{SEO}, this positive effect weakens or even turns negative, which indicates why the \textit{Pharma Good} coefficient is insignificant without considering co-occurring topics.

We can find further evidence for the influence of co-occurring topics on the announcement effects of news. The average effect of an announcement containing only the \textit{Debt} topic is negative with -1.27\%. However, if the acquired capital is used to repurchase stock, the overall effect turns positive.\\If \textit{Profit Warning} co-occurs with \textit{Guidance}, the overall effect decreases by more than 2.86 percentage points, as profit warnings on future earnings are likely to be more severe than profit warnings on past earnings.\\If \textit{Guidance} co-occurs with \textit{Bankruptcy Filing}, the overall effect increases by 9.22 percentage points. This might be interpreted as a situation where the company itself believes in its persistence by publishing earnings forecasts makes it more likely that this company will continue to exist. 

There would be further examples that emphasize the importance of co-occurring topics, all leading to the same finding. The multi-label structure allows a more fine-grained analysis of how topics influence abnormal returns. The impact of one topic on abnormal returns might vary significantly depending on the topics it co-occurs with.
\section{Conclusion}
\label{paper1_sec_conclusion}
This study adds to the financial topic model literature by providing the first topic model that is able to produce multi-label topic predictions. Furthermore, the model is based on BERT, which is why it benefits from state-of-the-art natural language processing techniques like extensive pre-training, self-attention and parallelism. The results yield evidence that the BERT based topic model significantly outperforms benchmarks like recurrent neural networks or standard feed-forward networks.

A further contribution of this paper is the creation of the ad-hoc multi-label database, which consists of 3,044 manually labeled German ad-hoc announcements that are assigned to 20 economically reasonable topics by financial experts.\\
The application of the topic model on stock market reactions reveals that stock markets reactions are highly topic dependent. Furthermore, stock market reactions on single topics also depend on the co-occurrence of other topics that appear in announcements, which confirms the contribution of the novel multi-label approach of the topic model derived in this work.

Nevertheless, the approach of this study has two major limitations which might be addressed in future research.  First, topics are gathered by screening ad-hoc announcements manually and identifying frequent and economically meaningful topics. Although the extracted topics are quite general and cover a wide range of contents in financial documents, there might be topics that are not covered yet. Hence, the topic model is not able to predict these latent topics. The greater the divergence between various financial texts and German ad-hoc announcements, the higher the likelihood of encountering topics in the respective documents that remain unaddressed. Therefore, future research might investigate whether these topics are also representative of the content of other data sources for financial text, as for example financial news, 10-K or 8-K filings. If they are not representative, one might add new topics to the list of my topics. However, this would induce the need of further manual labeling of text data and model retraining. Second, the results of the annotator performance of the ad-hoc multi-label database indicate that annotators tend to annotate too rarely rather than too often. This property translates directly to the model. Even though this problem is reduced by aggregating sentence level labeling to document level, future research might find different models or training ideas that address the imbalance of precision and recall in the ad-hoc multi-label database.

\newpage
\bibliographystyle{jf}
\bibliography{References.bib}
\newpage

\section*{Appendix}
\addcontentsline{toc}{section}{Appendix}

\subsection*{Annotator Instructions}
\label{paper1_sec_annotator_instructions}
The following instructions were handed to the annotators before the labeling phases. Prior to the first labeling phase, I conducted an online session with all annotators where I explained all topics in detail. After the first annotation phase, a second session was conducted where I focused on topics that have caused the most problems for the annotators. In this way, I minimized the probability of misunderstood topics.

\textit{General notes}: 
\begin{itemize}
\item Only label sentences that can be clearly assigned to a topic independently of previous and subsequent sentences in the message.
\begin{itemize}
\item \textit{Positive example}:\\"The member of the Management Board of 1st RED AG Jan German resigned from the Management Board on April 15, 2010."\\
$\rightarrow$ Management
\item \textit{Negative example}:\\"This is to be reviewed again at the end of 2015." 
\end{itemize}
However, it is sufficient if you are able to rule out all but one topic to assign a label.
\item Only current events are relevant. Historical events are not to be labeled, even if they fit thematically into a topic.
\end{itemize}
\textit{Topics}:
\begin{enumerate}
\item \textit{Earnings}: 
\begin{itemize}
\item Earnings Announcement, regular reporting on quarterly or annual results.
\item A sentence must be labeled if earnings figures are present (e.g. "Earnings per share (EPS) was EUR -0.37 in the first half of 2002 (previous year's value EUR -0.43) .")
\item If it is clear that the business result is reported in the announcement, it must also be annotated (e.g.: "Plaut Aktiengesellschaft, listed in the General Standard segment of the Frankfurt Stock Exchange, announces the Figures for the 1st half of 2014").
\item Explanations of how the results were generated or vague paraphrases are not to be annotated, as they cannot be clearly attributed to earnings.  (e.g.: "The improved situation was achieved with the help of the cost reduction program successfully implemented in 2002" or "Especially in Korea, the stabilization of the economic situation and the associated revival in demand is already noticeable").
\item Keywords: earnings, EBT, EBIT, EBITDA, R\&D, DB1, growth, increase, profit, loss, sales, (half) year, quarter, figures, announce, comparison, previous year.
\end{itemize}
\item \textit{Seasoned Equity Offering (SEO)}:
\begin{itemize}
\item Capital increase/reduction by issuing additional shares.
\item Keywords: capital increase, capital decrease, subscription right(phase), shares, subscribed, placed, increase, issue price, investors, acquire, note, note taking, volume.
\end{itemize}
\item \textit{Management}: 
\begin{itemize}
\item Any changes in management (board of directors, supervisory board, etc.).
\item Sentences with background information about manager/board are not relevant (e.g.: "Oliver Kaltner has worked for the Sony Group for the last five years").
\item Only sentences about executive board \& supervisory board are relevant, leading position is not sufficient (e.g.:" Peter Jansen was hired to lead the promotion team, an industry expert with many years of experience." $\rightarrow$ No management label).
\item If no name of the new/old CEO and/or no exact date of the management change is mentioned, no label is to be assigned (e.g.: "The finance department will be newly staffed as of November 1.").
\item Keywords: chairman of the board, CEO, appointed, office, member, supervisory board, resign, depart, successor.
\end{itemize}
\item \textit{Guidance}: 
\begin{itemize}
\item A company's forecast of its own profit or loss in the near future.
\item Expectations about past periods are not guidance!
\item Also label if it is clear that earnings/profit/loss forecasts are published in the report! (e.g.: "LifeWatch will integrate the business into existing operations to benefit from synergies (such as compensation and distribution ) and expects that the business can grow profitably in the coming years" is guidance!).
\item Statements about sales of the company not sufficient.
\item Keywords: forecast, expectation, result, profit, loss, EBIT, should, assume, expect.
\end{itemize}
\newpage
\item \textit{Profit Warning}:
\begin{itemize}
\item Surprising deterioration in earnings/earnings forecast.
\item Usually occurs in conjunction with Guidance or Earnings. Then please label both.
\item Keywords: loss, negative, reduce, impairment, EBIT.
\end{itemize}
\item \textit{M\&A}:
\begin{itemize}
\item New/expansion investment in company or own investment in other company, incl. acquisition.
\item Keywords: Takeover, acquires, purchase price, acquisition, shares, synergy, synergy effects.
\end{itemize}
\item \textit{Dividend}:
\begin{itemize}
\item Announcement dividend/dividend amount (incl. corrections and expectations).
\item Keywords: dividend, entitled to dividend, distribution, profits, entitled to profit, x euros per share.
\end{itemize}
\item \textit{Restructuring}: 
\begin{itemize}
\item Restructuring measures (processes, organization, capital structure, e.g.: Debt-equity swap, operational restructuring, etc.).
\item Usually occurs when the company is in crisis.
\item Separation of business unit/subsidiary is restructuring (e.g.: "Wuensche AG has sold its associated company, HL Hamburger Leistungsfutter GmbH \& Co. to a private investor, retroactive to January 1, 1997.").
\item Keywords: restructuring, restructure, reorganization, debt relief, operational, divestment, credit receivable reduced, secure financing, bridge financing.
\end{itemize}
\item \textit{Debt}:
\begin{itemize}
\item Company issues/returns loan/bond.
\item Keywords: (convertible, corporate) bond, (convertible) debenture, volume, loan, repayment, interest, interest rate, coupon, maturity, liabilities.
\end{itemize}
\newpage
\item \textit{Law}: 
\begin{itemize}
\item Company is involved in litigation, court case/investigation (case opened/closed, litigation accruals, sued).
\item Content of litigation is not relevant, only the litigation itself (e.g.: "DaimlerChrysler has agreed to settle a class action lawsuit pending in the United Stated District Court for the District of Delaware in connection with the 1998 business combination of Daimler-Benz and Chrysler to form DaimlerChrysler AG."$\rightarrow$ Law only, no M\&A).
\item Keywords: court, convicted, order, dismissed, prosecution, (investigative) proceedings, action, granted, appeal, objection, judgment, complaint, damages, authority review, litigation provisions.
\end{itemize}
\item \textit{Large Scale Project}:
\begin{itemize}
\item Completion of major project/order for the company.
\item Keywords: major order, contract, volume.
\end{itemize}
\item \textit{Squeeze Out}:
\begin{itemize}
\item Majority shareholder applies for squeeze (transfer of shares held by minority shareholders to majority shareholder), incl. progress of proceedings.
\item By definition only possible if at least 95\% of the share capital of a stock corporation is held.
\item Keywords: squeeze(-out), cash compensation, transfer of shares, remaining shareholders, majority shareholder, minority shareholder.
\end{itemize}
\item \textit{Bankruptcy Filing}:
\begin{itemize}
\item Company or third party has filed/will file for bankruptcy.
\item Bankruptcy application is new information .
\item In addition, information can be provided about contemplated future steps.
\item Keywords: bankruptcy application, bankruptcy, apply, insolvency, local court.
\end{itemize}
\item \textit{Bankruptcy Proceedings}:
\begin{itemize}
\item Information about concrete progress of bankruptcy proceedings is published.
\item Bankruptcy application is in the past.
\item Determined by the court, insolvency administrator, self-administration.
\item Keywords: bankruptcy plan, bankruptcy administrator, bankruptcy proceedings, self-administration.
\end{itemize}
\item \textit{Delay}:
\begin{itemize}
\item (Mandatory) report or annual general meeting is postponed or not published at all/does not take place.
\item If audit firm needs time $\rightarrow$ delay (e.g.: "Due to the extensive audit work caused by the expansion of the KIH Group's scope of consolidation and in view of the special investigations initiated by the new KIH Board of Management by the auditing firm C \& L Deutsche Revision AG, Duesseldorf  into the losses and responsibilities in detail, the date of the Annual General Meeting was set for August 18, 1998 as a precautionary measure.").
\item Keywords: delay, IFRS, cancel, postpone, annual financial statements.
\end{itemize}
\item \textit{Split}:
\begin{itemize}
\item Company carries out stock split.
\item Keywords: (share) split, ratio, new split, bonus share, additional share.
\end{itemize}
\item \textit{Pharma Good}:
\begin{itemize}
\item Drug approval/announcement/study success.
\item Keywords: (market) approval, FDA, study, results, treatment, drug, diagnosis, therapy, active ingredient, clinical, trial, application.
\end{itemize}
\item \textit{Buyback}:
\begin{itemize}
\item Repurchase of own shares.
\item Keywords: buyback, share buyback program, redeem, reduction of share capital.
\end{itemize}
\item \textit{Real Invest}:
\begin{itemize}
\item Buying or selling assets such as land, factories, machinery, etc.
\item Keywords: acquires, sells, build, office building, factory building, factory, land parcel, construction, production area, usable area, new, land, location.
\end{itemize}
\item \textit{Delisting}:
\begin{itemize}
\item Permanent delisting.
\item Keywords: delisting, revocation, shares, offer to purchase, termination of listing, stock exchange, terminate.
\end{itemize}
\item \textit{Irrelevant/Disclaimer}: 
\begin{itemize}
\item Sentence/section is not part of the core of the announcement (e.g. disclaimer, information about the company, information about the ad hoc obligation, etc.) or the announcement is in English.
\item Examples: "Prospectus is published...", "no investment recommendation", "listed on stock exchange" or also general company information that is independent of the announcement. (e.g." The INA Schaeffler Group is a global company for precision technology in the automotive, industrial and commercial sectors and employs around 54,000 people .", "If you have any questions, please call Joachim A. Klaehn at 040 - 36031 - 420!") .
\item Keywords: E-Mail, Freiverkehr, Prime Standard, will take place on, will be published on, will be offered in the following countries, further information, questions.
\end{itemize}
\end{enumerate}

\subsection*{Precision, Recall \& F1 BERT model}
\label{paper1_sec_precRecF1BertModel}
\begin{table}
\begin{flushleft}
\caption{Ad-Hoc Topic Model Precision, Recall \& F1}
 \mytablefont{This table shows the precision, recall and F1 scores of the BERT model for all topics as well as their micro and macro average (as defined in \citep{sokolova2009systematic}). I evaluate the performance on document level. I report the performance for models trained on sentence \& document level. Numbers are given in percent. Every model is trained for eight different seeds. I report the mean precision, recall and F1 score among seeds for every model and topic together with the standard deviation.}
\label{paper1_tab_model_precRecF1}
\mytablefont
\begin{tabularx}{\textwidth}{c *{8}{Y}}
\toprule
 & \multicolumn{4}{c}{\textit{Sentence Level}} & \multicolumn{4}{c}{\textit{Document Level}} \\
 & Num. & Precision & Recall & F1 & Num. & Precision & Recall & F1 \\
\midrule
Avg. (Macro) & 3,480 & \shortstack[c]{84.0 \\ (0.7)} & \shortstack[c]{87.3 \\ (0.7)} & \shortstack[c]{85.3 \\ (0.5)} & 1,029 & \shortstack[c]{90.3 \\ (0.8)} & \shortstack[c]{70.6 \\ (1.3)} & \shortstack[c]{78.6 \\ (0.9)} \\[0.3cm]
Avg. (Micro) & 3,480 & \shortstack[c]{82.9 \\ (0.8)} & \shortstack[c]{87.1 \\ (0.9)} & \shortstack[c]{85.0 \\ (0.6)} & 1,029 & \shortstack[c]{89.0 \\ (0.8)} & \shortstack[c]{70.3 \\ (1.3)} & \shortstack[c]{78.6 \\ (0.8)} \\ 
 \midrule
Squeeze Out & 152 & \shortstack[c]{93.2 \\ (0.9)} & \shortstack[c]{99.4 \\ (1.0)} & \shortstack[c]{96.2 \\ (0.6)} & 45 & \shortstack[c]{93.3 \\ (0)} & \shortstack[c]{93.3 \\ (0)} & \shortstack[c]{93.3 \\ (0)} \\[0.3cm]
Delay & 77 & \shortstack[c]{90.6 \\ (2.4)} & \shortstack[c]{97.3 \\ (1.1)} & \shortstack[c]{93.8 \\ (1.2)} & 33 & \shortstack[c]{95.8 \\ (3.7)} & \shortstack[c]{74.2 \\ (4.3)} & \shortstack[c]{83.5 \\ (2.9)} \\[0.3cm]
Delisting & 106 & \shortstack[c]{89.6 \\ (2.0)} & \shortstack[c]{95.4 \\ (2.1)} & \shortstack[c]{92.4 \\ (1.6)} & 35 & \shortstack[c]{91.0 \\ (2.7)} & \shortstack[c]{82.1 \\ (4.5)} & \shortstack[c]{86.3 \\ (2.1)} \\[0.3cm]
Management & 133 & \shortstack[c]{88.2 \\ (1.6)} & \shortstack[c]{96.6 \\ (1.7)} & \shortstack[c]{92.2 \\ (1.1)} & 52 & \shortstack[c]{92.0 \\ (2.6)} & \shortstack[c]{81.7 \\ (3.3)} & \shortstack[c]{86.5 \\ (1.6)} \\[0.3cm]
Law & 126 & \shortstack[c]{89.0 \\ (1.8)} & \shortstack[c]{93.4 \\ (2.0)} & \shortstack[c]{91.2 \\ (1.6)} & 38 & \shortstack[c]{99.2 \\ (1.5)} & \shortstack[c]{74.0 \\ (5.3)} & \shortstack[c]{84.6 \\ (3.3)} \\[0.3cm]
Dividend & 115 & \shortstack[c]{88.9 \\ (2.4)} & \shortstack[c]{93.5 \\ (0.9)} & \shortstack[c]{91.1 \\ (1.3)} & 50 & \shortstack[c]{87.3 \\ (1.9)} & \shortstack[c]{79.0 \\ (3.5)} & \shortstack[c]{82.9 \\ (2.5)} \\[0.3cm]
Debt & 161 & \shortstack[c]{85.4 \\ (3.1)} & \shortstack[c]{95.2 \\ (3.2)} & \shortstack[c]{90.0 \\ (2.6)} & 39 & \shortstack[c]{99.6 \\ (1.1)} & \shortstack[c]{72.8 \\ (3.6)} & \shortstack[c]{84.0 \\ (2.3)} \\[0.3cm]
Earnings & 914 & \shortstack[c]{84.4 \\ (1.0)} & \shortstack[c]{94.9 \\ (1.1)} & \shortstack[c]{89.4 \\ (0.8)} & 150 & \shortstack[c]{92.4 \\ (1.8)} & \shortstack[c]{83.9 \\ (1.6)} & \shortstack[c]{87.9 \\ (0.8)} \\[0.3cm]
Buyback & 132 & \shortstack[c]{88.3 \\ (5.6)} & \shortstack[c]{88.4 \\ (1.8)} & \shortstack[c]{88.2 \\ (2.7)} & 29 & \shortstack[c]{100.0 \\ (0)} & \shortstack[c]{79.7 \\ (2.9)} & \shortstack[c]{88.7 \\ (1.8)} \\[0.3cm]
Split & 86 & \shortstack[c]{84.6 \\ (4.0)} & \shortstack[c]{90.9 \\ (2.6)} & \shortstack[c]{87.6 \\ (2.1)} & 29 & \shortstack[c]{92.8 \\ (2.2)} & \shortstack[c]{72.0 \\ (2.9)} & \shortstack[c]{81.0 \\ (2.2)} \\[0.3cm]
Pharma Good & 97 & \shortstack[c]{84.4 \\ (5.4)} & \shortstack[c]{90.9 \\ (2.4)} & \shortstack[c]{87.5 \\ (3.4)} & 22 & \shortstack[c]{94.6 \\ (2.8)} & \shortstack[c]{78.4 \\ (2.1)} & \shortstack[c]{85.7 \\ (1.7)} \\[0.3cm]
Bankruptcy Filing & 118 & \shortstack[c]{80.9 \\ (2.3)} & \shortstack[c]{94.6 \\ (2.1)} & \shortstack[c]{87.2 \\ (1.6)} & 44 & \shortstack[c]{84.4 \\ (3.2)} & \shortstack[c]{75.0 \\ (2.4)} & \shortstack[c]{79.4 \\ (2.0)} \\[0.3cm]
Large Scale Project & 93 & \shortstack[c]{88.3 \\ (3.3)} & \shortstack[c]{85.4 \\ (2.4)} & \shortstack[c]{86.8 \\ (1.7)} & 35 & \shortstack[c]{99.6 \\ (1.1)} & \shortstack[c]{78.2 \\ (3.7)} & \shortstack[c]{87.6 \\ (1.9)} \\[0.3cm]
SEO & 198 & \shortstack[c]{76.4 \\ (5.3)} & \shortstack[c]{92.6 \\ (1.5)} & \shortstack[c]{83.6 \\ (3.1)} & 49 & \shortstack[c]{92.1 \\ (2.1)} & \shortstack[c]{65.1 \\ (6.0)} & \shortstack[c]{76.1 \\ (4.1)} \\[0.3cm]
Bankruptcy Proceedings & 125 & \shortstack[c]{80.2 \\ (2.5)} & \shortstack[c]{82.2 \\ (3.4)} & \shortstack[c]{81.2 \\ (2.2)} & 40 & \shortstack[c]{85.5 \\ (4.2)} & \shortstack[c]{69.7 \\ (5.1)} & \shortstack[c]{76.6 \\ (1.8)} \\[0.3cm]
Restructuring & 245 & \shortstack[c]{79.3 \\ (3.1)} & \shortstack[c]{80.2 \\ (2.3)} & \shortstack[c]{79.7 \\ (1.4)} & 97 & \shortstack[c]{86.0 \\ (2.8)} & \shortstack[c]{58.9 \\ (4.1)} & \shortstack[c]{69.8 \\ (3.0)} \\[0.3cm]
Guidance & 253 & \shortstack[c]{76.4 \\ (2.1)} & \shortstack[c]{81.5 \\ (2.6)} & \shortstack[c]{78.8 \\ (1.5)} & 108 & \shortstack[c]{77.0 \\ (1.7)} & \shortstack[c]{59.8 \\ (3.3)} & \shortstack[c]{67.3 \\ (1.8)} \\[0.3cm]
M \& A & 140 & \shortstack[c]{71.7 \\ (5.5)} & \shortstack[c]{72.7 \\ (4.1)} & \shortstack[c]{72.1 \\ (3.8)} & 49 & \shortstack[c]{82.6 \\ (3.7)} & \shortstack[c]{55.6 \\ (4.5)} & \shortstack[c]{66.4 \\ (4.0)} \\[0.3cm]
Real Invest & 72 & \shortstack[c]{85.0 \\ (6.7)} & \shortstack[c]{60.0 \\ (3.7)} & \shortstack[c]{70.2 \\ (3.3)} & 25 & \shortstack[c]{89.0 \\ (2.7)} & \shortstack[c]{36.5 \\ (5.8)} & \shortstack[c]{51.6 \\ (5.9)} \\[0.3cm]
Profit Warning & 137 & \shortstack[c]{75.8 \\ (5.1)} & \shortstack[c]{60.2 \\ (3.5)} & \shortstack[c]{67.0 \\ (2.5)} & 60 & \shortstack[c]{71.9 \\ (8.5)} & \shortstack[c]{41.0 \\ (5.5)} & \shortstack[c]{52.0 \\ (5.1)} \\
\bottomrule
\end{tabularx}
\end{flushleft}
\end{table}

Table \ref{paper1_tab_model_precRecF1} subdivides the F1 scores of the BERT model into precision and recall scores. We can see that on sentence level training, the average macro F1 score of 85.3\% is composed of a precision score of 84\% and a recall score of 87.3\%, which is a rather small gap between these two measures. Overall, one can conclude that the model produces robust topic predictions on average. However, the slightly better recall score indicates that the model tends to label too often, which tends to decrease the number of false negatives at the cost of an increased number of false positives. However, this result is a direct consequence of Tables \ref{paper1_prec_rec_f1_category_1} and \ref{paper1_prec_rec_f1_category_2}, which present the annotator performance per topic for the respective labeling rounds. Due to a higher precision than recall, these tables provide evidence that the annotators produce high quality labels, but they labeled too rarely. Assuming that the BERT model learns the topic specific patterns well due to the high quality topic labels, the model also correctly labels the sentences that were left blank by mistake by the annotators. That leads to more false positively labeled sentences with respect to annotator gold labels. The higher the discrepancy between precision and recall in the gold labels, the higher the respective discrepancy in the model predictions, only with reversed sign (for example the \textit{SEO} topic). On document level, this pattern is reversed.  
\subsection*{Performance English Models}
\label{paper1_sec_EnglishBert}
Most of financial text is given in English. Therefore, after translating the ad-hoc multi-label database, I train the BERT model and all benchmarks with exactly the same settings as in the German case, with one exception: For the BERT-base model, I use the uncased version, as cased words only play a minor role for English texts.

\begin{table}
\begin{flushleft}
\caption{Ad-Hoc Topic Model Performance (English Data)}
 \mytablefont{This table compares the multi-label classification performance of the uncased version of the English BERT base model with an ELMo model with GRU layers and pre-trained, fasttext word embeddings \citep{ELMO}, called \textit{BiGRU}, a standard GRU model \citep{GRU} and a two-layer feed-forward neural network (called \textit{NN}). The latter two models use randomly initialized word embeddings. Every model is trained for eight different seeds. I report the mean F1 score among seeds for every model and topic together with the standard deviation. In addition, I report the micro and macro F1 scores \citep{sokolova2009systematic} as overall performance measures and the number of labels per topic. I evaluate the performance on document level. I report the performance of models that are trained on sentence \& document level. Numbers are given in percent.}
\label{paper1_tab_model_performance_en}
\mytablefont
\begin{tabularx}{\textwidth}{c *{10}{Y}}
\toprule
 & \multicolumn{5}{c}{\textit{Sentence Level}} & \multicolumn{5}{c}{\textit{Document Level}} \\
 & Num. & BERT & BiGRU & GRU & NN & Num. & BERT & BiGRU & GRU & NN \\
\midrule
Avg. (Macro) & 3,480 & \shortstack[c]{84.7 \\ (0.5)} & \shortstack[c]{79.0 \\ (0.8)} & \shortstack[c]{53.0 \\ (7.3)} & \shortstack[c]{67.0 \\ (0.7)} & 1,029 & \shortstack[c]{75.8 \\ (0.7)} & \shortstack[c]{73.6 \\ (0.5)} & \shortstack[c]{0.1 \\ (0.2)} & \shortstack[c]{8.9 \\ (1.1)} \\[0.25cm]
Avg. (Micro) & 3,480 & \shortstack[c]{84.3 \\ (0.5)} & \shortstack[c]{80.6 \\ (0.6)} & \shortstack[c]{64.4 \\ (5.0)} & \shortstack[c]{72.8 \\ (0.6)} & 1,029 & \shortstack[c]{77.3 \\ (0.4)} & \shortstack[c]{74.3 \\ (0.4)} & \shortstack[c]{0.3 \\ (0.5)} & \shortstack[c]{26.0 \\ (1.6)} \\ 
 \midrule
Delisting & 106 & \shortstack[c]{95.5 \\ (1.2)} & \shortstack[c]{90.8 \\ (1.8)} & \shortstack[c]{75.1 \\ (9.5)} & \shortstack[c]{82.7 \\ (3.9)} & 35 & \shortstack[c]{87.8 \\ (2.8)} & \shortstack[c]{87.6 \\ (2.1)} & \shortstack[c]{0 \\ (0)} & \shortstack[c]{0 \\ (0)} \\[0.25cm]
Squeeze Out & 152 & \shortstack[c]{95.2 \\ (0.8)} & \shortstack[c]{94.4 \\ (1.1)} & \shortstack[c]{93.5 \\ (1.4)} & \shortstack[c]{90.5 \\ (2.1)} & 45 & \shortstack[c]{93.6 \\ (1.4)} & \shortstack[c]{92.5 \\ (1.4)} & \shortstack[c]{0 \\ (0)} & \shortstack[c]{60.0 \\ (20.8)} \\[0.25cm]
Management & 133 & \shortstack[c]{94.0 \\ (1.0)} & \shortstack[c]{88.6 \\ (1.6)} & \shortstack[c]{86.5 \\ (2.6)} & \shortstack[c]{89.9 \\ (1.4)} & 52 & \shortstack[c]{87.0 \\ (1.2)} & \shortstack[c]{80.5 \\ (3.3)} & \shortstack[c]{0 \\ (0)} & \shortstack[c]{0 \\ (0)} \\[0.25cm]
Delay & 77 & \shortstack[c]{92.9 \\ (1.3)} & \shortstack[c]{86.9 \\ (2.6)} & \shortstack[c]{63.2 \\ (27.2)} & \shortstack[c]{72.2 \\ (3.1)} & 33 & \shortstack[c]{82.2 \\ (3.2)} & \shortstack[c]{76.6 \\ (4.1)} & \shortstack[c]{0 \\ (0)} & \shortstack[c]{0 \\ (0)} \\[0.25cm]
Pharma Good & 97 & \shortstack[c]{91.0 \\ (2.4)} & \shortstack[c]{88.9 \\ (4.6)} & \shortstack[c]{20.9 \\ (23.7)} & \shortstack[c]{87.2 \\ (5.4)} & 22 & \shortstack[c]{90.0 \\ (3.7)} & \shortstack[c]{87.8 \\ (4.8)} & \shortstack[c]{0 \\ (0)} & \shortstack[c]{0 \\ (0)} \\[0.25cm]
Dividend & 115 & \shortstack[c]{89.7 \\ (0.6)} & \shortstack[c]{91.6 \\ (1.4)} & \shortstack[c]{87.1 \\ (1.5)} & \shortstack[c]{89.5 \\ (0.5)} & 50 & \shortstack[c]{85.5 \\ (2.4)} & \shortstack[c]{80.4 \\ (1.9)} & \shortstack[c]{0 \\ (0)} & \shortstack[c]{1.0 \\ (2.7)} \\[0.25cm]
Earnings & 914 & \shortstack[c]{89.4 \\ (0.7)} & \shortstack[c]{88.5 \\ (0.8)} & \shortstack[c]{86.4 \\ (0.6)} & \shortstack[c]{85.9 \\ (1.1)} & 150 & \shortstack[c]{89.1 \\ (0.9)} & \shortstack[c]{82.6 \\ (1.1)} & \shortstack[c]{1.9 \\ (3.2)} & \shortstack[c]{77.0 \\ (2.0)} \\[0.25cm]
Law & 126 & \shortstack[c]{89.2 \\ (2.0)} & \shortstack[c]{85.8 \\ (3.7)} & \shortstack[c]{58.1 \\ (23.6)} & \shortstack[c]{73.4 \\ (4.8)} & 38 & \shortstack[c]{84.6 \\ (3.4)} & \shortstack[c]{85.3 \\ (1.6)} & \shortstack[c]{0 \\ (0)} & \shortstack[c]{0 \\ (0)} \\[0.25cm]
Split & 86 & \shortstack[c]{87.8 \\ (2.6)} & \shortstack[c]{84.9 \\ (2.5)} & \shortstack[c]{64.4 \\ (18.6)} & \shortstack[c]{79.8 \\ (2.1)} & 29 & \shortstack[c]{78.7 \\ (2.6)} & \shortstack[c]{82.2 \\ (3.0)} & \shortstack[c]{0 \\ (0)} & \shortstack[c]{0 \\ (0)} \\[0.25cm]
Large Scale Project & 93 & \shortstack[c]{86.9 \\ (1.5)} & \shortstack[c]{83.0 \\ (3.7)} & \shortstack[c]{4.3 \\ (12.0)} & \shortstack[c]{80.4 \\ (3.9)} & 35 & \shortstack[c]{84.3 \\ (3.2)} & \shortstack[c]{80.4 \\ (2.4)} & \shortstack[c]{0 \\ (0)} & \shortstack[c]{0 \\ (0)} \\[0.25cm]
Buyback & 132 & \shortstack[c]{86.8 \\ (1.6)} & \shortstack[c]{87.8 \\ (2.5)} & \shortstack[c]{59.6 \\ (36.9)} & \shortstack[c]{85.2 \\ (3.3)} & 29 & \shortstack[c]{84.0 \\ (2.3)} & \shortstack[c]{84.2 \\ (2.3)} & \shortstack[c]{0 \\ (0)} & \shortstack[c]{0 \\ (0)} \\[0.25cm]
Debt & 161 & \shortstack[c]{85.5 \\ (1.7)} & \shortstack[c]{81.7 \\ (3.1)} & \shortstack[c]{76.3 \\ (4.1)} & \shortstack[c]{74.5 \\ (0.7)} & 39 & \shortstack[c]{82.0 \\ (1.9)} & \shortstack[c]{81.9 \\ (2.5)} & \shortstack[c]{0 \\ (0)} & \shortstack[c]{0 \\ (0)} \\[0.25cm]
SEO & 198 & \shortstack[c]{85.4 \\ (2.4)} & \shortstack[c]{82.2 \\ (2.2)} & \shortstack[c]{67.9 \\ (13.7)} & \shortstack[c]{80.3 \\ (4.2)} & 49 & \shortstack[c]{72.8 \\ (4.0)} & \shortstack[c]{72.7 \\ (4.9)} & \shortstack[c]{0 \\ (0)} & \shortstack[c]{0 \\ (0)} \\[0.25cm]
Bankruptcy Filing & 118 & \shortstack[c]{85.2 \\ (1.8)} & \shortstack[c]{80.6 \\ (4.0)} & \shortstack[c]{67.6 \\ (4.1)} & \shortstack[c]{70.9 \\ (1.8)} & 44 & \shortstack[c]{77.3 \\ (2.8)} & \shortstack[c]{76.2 \\ (3.6)} & \shortstack[c]{0 \\ (0)} & \shortstack[c]{0 \\ (0)} \\[0.25cm]
Restructuring & 245 & \shortstack[c]{79.3 \\ (0.8)} & \shortstack[c]{72.8 \\ (4.0)} & \shortstack[c]{24.0 \\ (26.2)} & \shortstack[c]{61.9 \\ (5.5)} & 97 & \shortstack[c]{68.5 \\ (2.7)} & \shortstack[c]{62.5 \\ (3.6)} & \shortstack[c]{0 \\ (0)} & \shortstack[c]{0 \\ (0)} \\[0.25cm]
Bankruptcy Proceedings & 125 & \shortstack[c]{77.5 \\ (1.4)} & \shortstack[c]{75.6 \\ (3.4)} & \shortstack[c]{67.9 \\ (14.5)} & \shortstack[c]{70.2 \\ (5.3)} & 40 & \shortstack[c]{66.5 \\ (4.6)} & \shortstack[c]{74.9 \\ (3.2)} & \shortstack[c]{0 \\ (0)} & \shortstack[c]{0 \\ (0)} \\[0.25cm]
Guidance & 253 & \shortstack[c]{77.1 \\ (1.8)} & \shortstack[c]{76.1 \\ (1.1)} & \shortstack[c]{58.0 \\ (7.8)} & \shortstack[c]{66.1 \\ (1.9)} & 108 & \shortstack[c]{65.1 \\ (1.7)} & \shortstack[c]{64.4 \\ (2.3)} & \shortstack[c]{0.5 \\ (0.8)} & \shortstack[c]{39.4 \\ (3.4)} \\[0.25cm]
Real Invest & 72 & \shortstack[c]{72.8 \\ (4.4)} & \shortstack[c]{23.7 \\ (9.7)} & \shortstack[c]{0 \\ (0)} & \shortstack[c]{0 \\ (0)} & 25 & \shortstack[c]{19.3 \\ (12.7)} & \shortstack[c]{12.5 \\ (8.0)} & \shortstack[c]{0 \\ (0)} & \shortstack[c]{0 \\ (0)} \\[0.25cm]
M \& A & 140 & \shortstack[c]{70.0 \\ (2.6)} & \shortstack[c]{66.8 \\ (3.9)} & \shortstack[c]{0 \\ (0)} & \shortstack[c]{0 \\ (0)} & 49 & \shortstack[c]{63.9 \\ (4.9)} & \shortstack[c]{55.2 \\ (4.9)} & \shortstack[c]{0 \\ (0)} & \shortstack[c]{0 \\ (0)} \\[0.25cm]
Profit Warning & 137 & \shortstack[c]{63.4 \\ (2.1)} & \shortstack[c]{49.1 \\ (4.9)} & \shortstack[c]{0 \\ (0)} & \shortstack[c]{0 \\ (0)} & 60 & \shortstack[c]{53.1 \\ (4.0)} & \shortstack[c]{51.9 \\ (3.2)} & \shortstack[c]{0 \\ (0)} & \shortstack[c]{0 \\ (0)} \\
\bottomrule
\end{tabularx}
\end{flushleft}
\end{table}

As we can see in Table \ref{paper1_tab_model_performance_en}, the BERT model achieves a similar performance as its German counterpart with an average macro F1 score of 84.7\%. The BERT model strongly outperforms its benchmarks also in case of English data. All findings for German training data repeat for English data.

\end{document}